\documentclass[12pt]{article}
\usepackage{graphics}
\usepackage{amssymb}
\usepackage{amsmath}
\def\preprint%
#1{\thispagestyle{empty}~\newline\vspace*{-22.65mm}%
\begin{flushright}%
\begin{tabular}{l} #1%
\end{tabular}%
\end{flushright}%
\vspace{1cm}}%
\renewcommand{\theequation}{\arabic{section}.\arabic{equation}}
\renewcommand{\thesubsection}{\arabic{section}.\arabic{subsection}}
\begin{document}
\markright{\hfil Foundations of anisotropic relativistic mechanics}
\title{\bf \LARGE Foundations of anisotropic relativistic mechanics}
\author{Sebastiano Sonego\thanks{\tt
sebastiano.sonego@uniud.it}\hspace{2mm} 
and 
Massimo Pin\thanks{\tt pin@fisica.uniud.it}\\[2mm] 
{\small \it Universit\`a di Udine, Via delle Scienze 208,
33100 Udine, Italy}}

\date{{\small June 30, 2009; \LaTeX-ed \today}}
\maketitle
\begin{abstract}
We lay down the foundations of particle dynamics in mechanical theories that satisfy the relativity principle and whose kinematics can be formulated employing reference frames of the type usually adopted in special relativity.  Such mechanics allow for the presence of anisotropy, both conventional (due to non-standard synchronisation protocols) and real (leading to detectable chronogeometrical effects, independent of the choice of synchronisation).  We give a general method for finding the fundamental dynamical quantities (Lagrangian, energy and momentum)  and write their explicit expression in all the kinematics compatible with the basic requirements.  We also write the corresponding dispersion relations and outline a formulation of these theories in terms of a pseudo-Finslerian spacetime geometry.  Although the treatment is restricted to the case of one spatial dimension, an extension to three dimensions is almost straightforward.\\

\end{abstract}
\def\ee{{\mathrm e}}
\def\d{{\mathrm d}}
\def\g{\mbox{\sl g}}
\def\SIZE{1.00}

\newpage
\section{Introduction}
\label{sec:intro}

The goal of the present paper is to lay down the foundations of dynamics for mechanical theories satisfying the principle of relativity.

We have shown elsewhere~\cite{sp,sp-3d} that, starting from the composition law for velocities and using the principle of relativity, the usual definition of kinetic energy for a particle (as a scalar quantity whose change equals the work done on the particle), and assuming the existence of elastic collisions between asymptotically free particles, one can construct such quantities as momentum, kinetic energy, the Lagrangian and the Hamiltonian for a free particle in an inertial frame --- that is, all the basic ingredients one needs in order to build up dynamics.  Although the treatment in Ref.~\cite{sp} was restricted only to the pedagogically relevant cases of Newtonian and Einstein's mechanics, the basic technique is general and can be applied to produce a wider class of theories --- essentially, all those in which the principle of relativity holds.

How general are such theories?  Galilei's and Einstein's kinematics require homogeneity of space and time, isotropy of space, the relativity principle, and a pre-causality condition (see Sec.~\ref{subsec:trans} below for more details).\footnote{The existence of a (possibly infinite) invariant speed {\em follows\/} from these hypotheses and does not require an independent postulate.  The actual value of such a speed is an experimental issue.}  At first then, one might expect that one could generalise mechanics a good deal, by relaxing one or more of these hypotheses.  However, on closer inspection it turns out that there is not much room left.  Indeed, homogeneity is crucial in order to set up the very notion of a reference frame of the type commonly considered in special relativity.  Moreover, as we wrote at the beginning, we are interested in theories that do satisfy the relativity principle.  And, finally, pre-causality is fundamental if we want to be able to do physics at all. Hence, the only hypothesis we can drop is the one of isotropy.  By  this, we do not mean {\em geometric\/} isotropy (we still assume that space be Euclidean in every inertial frame) but the different notion of  {\em mechanical\/} isotropy --- that all oriented directions in space are equivalent in kinematics and free particle dynamics.

The most general kinematics satisfying all the other postulates except mechanical isotropy were studied by Lalan in 1937~\cite{lalan}, and will be reviewed and commented in Sec.~\ref{sec:kin} in order to provide a self-contained presentation.  In these kinematics, anisotropy manifests itself in two ways.  First of all, there could be different invariant speeds along different directions.  Thus, if we assume that light in vacuum travels at the invariant speed, as usual,\footnote{There is no fundamental reason why it should be so~\cite{light}, but there is excellent experimental evidence that any difference is very small. This justifies our identification.} it follows that in these theories it does not propagate isotropically.  Furthermore, the factor that relates the measurements of time intervals by clocks in relative motion, and the analogous factor for measurements of lengths,  depend on the direction of the relative velocity.  Interestingly, these two manifestations of mechanical anisotropy are independent of each other --- a circumstance which should not be entirely surprising, because the first type of effect is an artifact due to a non-standard convention for synchronisation of distant clocks, whereas the second one is objective and cannot be gauged away.  In Sec.~\ref{sec:general} we describe a  general method for constructing the fundamental dynamical quantities corresponding to a given relativistic kinematics, which is then applied systematically in Secs.~\ref{sec:examples} and \ref{sec:hamiltonian} to the cases classified in Sec.~\ref{sec:kin}.  The possibility for giving these theories a geometrical formulation in a four-dimensional spacetime is discussed in Sec.~\ref{sec:geometry}.  Section~\ref{sec:comments} contains a few additional comments.  

We restrict ourselves to considering motion along one space dimension.  The extension to three space dimensions is left for future investigations; see also Ref.~\cite{sp-3d}.

\section{Anisotropic relativistic kinematics}
\label{sec:kin}
\setcounter{equation}{0}

We begin developing systematically the kinematics compatible with the principle of relativity.  In Sec.~\ref{subsec:trans} we limit ourselves to derive and classify the transformation laws (see also Ref.~\cite{lalan}), leaving all comments and remarks to Sec.~\ref{subsec:interpretation}.

\subsection{Transformation law}
\label{subsec:trans}

Consider a reference frame $\cal K$ that moves with constant velocity with respect to another reference frame $\overline{\cal K}$, and assume that the corresponding temporal and spatial coordinates\footnote{It is perhaps worth stating explicitly that the ``coordinates'' used in this paper (and in almost all the literature on special relativity) correspond to well-defined readings of time and distance, so they are not just arbitrary labels, but have a clear operational meaning~\cite{ws}.  Thus, all statements about the vague notions of ``time'' and ``space'' can be unambiguously interpreted in terms of the behaviour of clocks and rulers (or other physical systems  used in measurement protocols).} $(t,x)$ and $(\bar{t},\bar{x})$ are chosen in such a way that $t=x=0$ iff $\bar{t}=\bar{x}=0$.  Then, the most general coordinate transformation between the systems $\cal K$ and $\overline{\cal K}$ that is compatible with temporal and spatial homogeneity is~\cite{lalan,eisenberg}
\begin{equation}
\left.\begin{array}{l}
\bar{t}=A(v)\,t+B(v)\,x\\
\bar{x}=C(v)\,t+D(v)\,x
\end{array}\right\}\;,
\label{trans}
\end{equation}
where $A$, $B$, $C$ and $D$ are arbitrary functions of $v$, the velocity\footnote{In order for the notion of velocity to make sense operationally, it is obvious that {\em some\/} prescription must have  been adopted for the synchronisation of clocks in a reference frame.  We do not need to specify what the prescription is --- we only require it to be compatible with the relativity principle.} of $\cal K$ with respect to $\overline{\cal K}$.  The parameter $v$ belongs to some open interval $J\subseteq\mathbb{R}$, containing $0$. Hereafter, we shall write $J=:\,]-c_-,c_+\,[\,$, with $c_-$ and $c_+$ positive constants (possibly infinite).

The condition that $x=0$ iff $\bar{x}=v\bar{t}$, implies $C(v)=A(v)v$, so we can rewrite Eq.~(\ref{trans}) as
\begin{equation}
\left(\begin{array}{c}
\bar{t}\\ \bar{x}
\end{array}\right)=\Lambda(v)
\left(\begin{array}{c}
t\\ x
\end{array}\right)\;,
\label{trans-matrix}
\end{equation}
where $\Lambda(v)$ is the matrix
\begin{equation}
\Lambda(v):=A(v)
\left(\begin{array}{cc}
1 & \xi(v) \\
v & \eta(v)
\end{array}\right)\;,
\label{lambda}
\end{equation}
and we have introduced the ratios $\xi(v):=B(v)/A(v)$ and $\eta(v):=D(v)/A(v)$. Obviously, $\Lambda(0)$ must be the identity matrix, so $A(0)=\eta(0)=1$ and $\xi(0)=0$.

It has been argued by several authors that the relativity principle can be implemented at the kinematical level, by requiring that the transformation (\ref{trans-matrix}) be associative.  Let us consider also a third reference frame $\overline{\overline{\cal K}}$, which moves with constant velocity $u$ with respect to $\cal K$.  The velocity $\bar{u}$ of $\overline{\overline{\cal K}}$ with respect to $\overline{\cal K}$ will be given by some composition law\footnote{By the relativity principle, the velocities will all belong to the same open interval $J\subseteq\mathbb{R}$, independently of the reference frame in which they are measured, so $\Phi:J\times J\to J$.}
\begin{equation}
\bar{u}=\Phi(u,v)\;.
\label{comp-gen}
\end{equation}
The transformation between $\overline{\cal K}$ and $\overline{\overline{\cal K}}$ is
\begin{equation}
\left(\begin{array}{c}
\bar{t}\\ \bar{x}
\end{array}\right)=\Lambda(\bar{u})
\left(\begin{array}{c}
\bar{\!\bar{t}}\\ \bar{\bar{x}}
\end{array}\right)\;,
\label{trans-matrix'}
\end{equation}
so the transformation between $\cal K$ and $\overline{\overline{\cal K}}$ is
\begin{equation}
\left(\begin{array}{c}
t\\ x
\end{array}\right)=\Lambda(v)^{-1}\Lambda(\bar{u})
\left(\begin{array}{c}
\bar{\!\bar{t}}\\ \bar{\bar{x}}
\end{array}\right)\;.
\label{trans-matrix''}
\end{equation}
By the relativity principle, this transformation must be of the same type as those that link $\cal K$ and $\overline{\overline{\cal K}}$ to $\overline{\cal K}$, so it must be possible to replace $\Lambda(v)^{-1}\Lambda(\bar{u})$ by $\Lambda(u)$, which amounts to
\begin{equation}
\Lambda(v)\Lambda(u)=\Lambda(\Phi(u,v))\;.
\label{associative}
\end{equation}

As we shall see in a moment, this requirement allows one to determine the functions $A$, $\xi$ and $\eta$ in terms of\footnote{Hereafter, a prime will denote the derivative of a function with respect to its argument, with only a few obvious exceptions --- see, {\em e.g.\/}, Eq.~(\ref{integral}) below.} $A'(0)$, $\xi'(0)$, and $\eta'(0)$.  Note that, defining as $v^\ast$ the velocity of a frame with respect to $\cal K$ such that $\Phi(v^\ast,v)=0$, and remembering that $\Lambda(0)$ is the identity matrix $I$, Eq.~(\ref{associative}) implies $\Lambda(v)\Lambda(v^\ast)=I$.  Hence, imposing the relativity principle is tantamount to giving group structure to the set of coordinate transformations.

Differentiating Eq.~(\ref{associative}) with respect to $v$ and evaluating the result for $v=0$ we find
\begin{equation}
\Lambda'(0)\Lambda(u)=\varphi(u)\Lambda'(u)\;,
\label{eqforLambda}
\end{equation}
where we have defined the function
\begin{equation}
\varphi(u):=\left.\frac{\partial \Phi(u,v)}{\partial
v}\right|_{v=0}\;.
\label{f(u)}
\end{equation}
Equation (\ref{eqforLambda}) is a first-order, linear differential equation for $\Lambda$, whose solution is
\begin{equation}
\Lambda(u)=\exp\left(h(u)\Lambda'(0)\right)\;,
\label{expLambda}
\end{equation}
where we have used the condition $\Lambda(0)=I$, and
\begin{equation}
h(u):=\int_0^u\frac{\d u'}{\varphi(u')}\;.
\label{integral}
\end{equation}

The structure of $\Lambda'(0)$ is obtained directly from Eq.\ (\ref{lambda}).  One finds
\begin{equation}
\Lambda'(0)=\frac{1}{\kappa}\,I+M\;,
\label{I+M}
\end{equation}
where
\begin{equation}
M:=\left(\begin{array}{cc}
-b/2 & a \\
1 & b/2
\end{array}\right)\;,
\label{M}
\end{equation}
and $a:=\xi'(0)$, $b:=\eta'(0)$, $1/\kappa:=A'(0)+b/2$.  In particular, note that $\kappa$ is, dimensionally, a velocity; it can have either sign and can also be infinite.

The matrix $M$ has the following remarkable properties: For any natural number $k$,
\begin{equation}
M^{2k}=\left(a+b^2/4\right)^k I\;,\qquad\qquad
M^{2k+1}=\left(a+b^2/4\right)^k M\;.
\label{propM}
\end{equation}
Replacing these into Eq.~(\ref{expLambda}) we find
\begin{equation}
\Lambda(u)=\ee^{h(u)/\kappa}
\left(\begin{array}{cc}
s_1(u)-b\,s_2(u)/2 & a\,s_2(u) \\
s_2(u) & s_1(u)+b\,s_2(u)/2
\end{array}\right)\;,
\label{Lambda!}
\end{equation}
where:
\begin{equation}
s_1(u):=\sum_{k=0}^{+\infty}\frac{1}{(2k)!}\left(a+b^2/4\right)^k
h(u)^{2k}\;;
\label{sigma1}
\end{equation}
\begin{equation}
s_2(u):=\sum_{k=0}^{+\infty}\frac{1}{(2k+1)!}\left(a+b^2/4\right)^k
h(u)^{2k+1}\;.
\label{sigma2}
\end{equation}
Interestingly, $s_1$ and $s_2$ satisfy the identity
\begin{equation}
s_1(u)^2-\left(a+b^2/4\right)s_2(u)^2=1\;,
\label{idsigma}
\end{equation}
so $\Lambda(u)$ is, by Eq.~(\ref{Lambda!}), just equal to $\ee^{h(u)/\kappa}$ times a unimodular matrix.

Comparing Eqs.~(\ref{lambda}) and (\ref{Lambda!}) we find:
\begin{equation}
A(u)=\ee^{h(u)/\kappa}\left(s_1(u)-b\,s_2(u)/2\right)\;;
\label{A}
\end{equation}
\begin{equation}
u=\left(\frac{s_1(u)}{s_2(u)}-\frac{b}{2}\right)^{-1}\;;
\label{u}
\end{equation}
\begin{equation}
\xi(u)=a\,u\;;
\label{xi}
\end{equation}
\begin{equation}
\eta(u)=1+b\,u\;.
\label{eta}
\end{equation}
Note that Eq.~(\ref{u}) contains implicitly the link between $u$ and $h(u)$, because its right-hand side depends on $u$ only through $h(u)$.

Replacing Eq.~(\ref{Lambda!}) and the corresponding expressions for $\Lambda(v)$ and $\Lambda(\Phi(u,v))$ into Eq.~(\ref{associative}) we find:
\begin{equation}
h(\Phi(u,v))=h(u)+h(v)\;;
\label{h+h}
\end{equation}
\begin{equation}
s_1(\Phi(u,v))=s_1(u)s_1(v)
+\left(a+b^2/4\right)s_2(u)s_2(v)\;;
\label{sigma1Phi}
\end{equation}
\begin{equation}
s_2(\Phi(u,v))=s_1(u)s_2(v)+s_2(u)s_1(v)\;.
\label{sigma2Phi}
\end{equation}
Equation (\ref{h+h}) tells us that $U:=h(u)$ is additive --- the canonical parametrisation for the group~\cite{lalan}.  Equations (\ref{sigma1Phi}) and (\ref{sigma2Phi}) allow one to find the composition law between $u$ and $v$, using repeatedly Eq.\ (\ref{u}):
\begin{equation}
\Phi(u,v)=\left(\frac{s_1(\Phi(u,v))}{s_2(\Phi(u,v))}
-\frac{b}{2}\right)^{-1}=\frac{u+v+b\,u\,v}{1+a\,u\,v}\;.
\label{Phiab}
\end{equation}
This corresponds to
\begin{equation}
\varphi(u)=-a\,u^2+b\,u+1\;.
\label{phiab'}
\end{equation}
Finally, using the relationship $s_2'(u)=s_1(u)/\varphi(u)$, which follows from Eqs.~(\ref{integral}), (\ref{sigma1}) and (\ref{sigma2}), together with Eqs.~(\ref{u}) and (\ref{phiab'}), we obtain
\begin{equation}
\frac{s_2'(u)}{s_2(u)}=\left(\frac{1}{u}
+\frac{b}{2}\right)\frac{1}{-a\,u^2+b\,u+1}=\frac{1}{u}
-\frac{\varphi'(u)}{2\,\varphi(u)}\;.
\label{sigma'/sigma}
\end{equation}
This can be immediately integrated to obtain, in a neighbourhood of the origin, $s_2(u)=u\,\varphi(u)^{-1/2}$; hence
\begin{equation}
A(u)=\ee^{h(u)/\kappa}\,\varphi(u)^{-1/2}\;.
\label{A!}
\end{equation}
An alternative derivation of all these results is presented in Appendix~\ref{subsec:appendixA}.

Equation (\ref{A!}) provides one with a criterion for finding the limit velocities $-c_-$ and $c_+$.  Since $\varphi(0)=1$, it follows that $J$ is the largest interval containing $0$ for which $\varphi(u)>0$.  The finite values of the limit velocities can then be found solving the equation $\varphi(u)=0$.  It is not difficult to check explicitly that if $u,v\in J$, then also $\Phi(u,v)\in J$, as expected for consistency.

All these results are the most general ones compatible only with homogeneity and with the relativity principle.  As already pointed out by Lalan~\cite{lalan} and others, however, not all the possible values for $a$ and $b$ are suitable for describing possible kinematics.  Indeed, imposing also a natural pre-causality condition, that two events happening at the same place in a reference frame must be in the same causal relationship in any other frame, it follows that $\partial \bar{t}/\partial t>0$, {\em i.e.\/}, that $A(u)>0$, $\forall u\in J$.  This is equivalent to
\begin{equation}
s_1(u)-b\,s_2(u)/2>0\;,\qquad\forall u\in J\;.
\label{precausality}
\end{equation}

In order to proceed further and see what kind of constraints the inequality (\ref{precausality}) imposes on $a$ and $b$, we need to enter a tedious case-by-case analysis.  Before doing so, however, it is convenient to change notation and use, in place of $a$ and $b$, the three parameters $c>0$, $\sigma\in\{1,-1,0\}$, and $\varepsilon\in\mathbb{R}$ defined through the relations
\begin{equation}
a=\left(\sigma-\varepsilon^2\right)/c^2\;,\qquad\qquad\qquad
b=2\,\varepsilon/c\;,
\label{ab}
\end{equation}
such that $a+b^2/4=\sigma/c^2$.  Expressed in terms of these new parameters, the limit velocities are
\begin{equation}
-\frac{c}{\sqrt{\sigma}+\varepsilon}\qquad\mbox{and}\qquad
\frac{c}{\sqrt{\sigma}-\varepsilon}\;.
\label{cepsilon}
\end{equation}
Of course, no limit velocities exist (not even with infinite value) for $\sigma=-1$.

\subsubsection{Case $\sigma=1$, $c<+\infty$}
\label{subsubsec:1}

We have
\begin{equation}
s_1(v)=\cosh\left(h(v)/c\right)\;,\qquad\qquad\qquad
s_2(v)=c\,\sinh\left(h(v)/c\right)\;,
\label{ss>0}
\end{equation}
so the pre-causality condition (\ref{precausality}) is satisfied only for $|\varepsilon|\leq 1$ (which is equivalent to $a\geq 0$).\footnote{Also, for $\varepsilon>1$ the limit velocities in (\ref{cepsilon}) are both negative, and for $\varepsilon<-1$ they are both positive.  This violates the condition $0\in J$.}

For $|\varepsilon|<1$, the equation $\varphi(v)=0$ has two distinct roots $c_+$ and $-c_-$ of opposite sign, with $c_\pm=c/\left(1\mp\varepsilon\right)$, and Eq.~(\ref{u}) gives
\begin{equation}
h(v)=c\,\ln\left(\frac{1+\left(1+\varepsilon\right)v/c}{1
-\left(1-\varepsilon\right)v/c}\right)^{1/2}\;,
\label{h-case1}
\end{equation}
so $h$ maps $J=\,]-c_-,c_+\,[\,$ onto $\mathbb{R}$.  The transformation is therefore
\begin{equation}
\left.\begin{array}{l}
\bar{t}={\displaystyle
\left(\frac{1+\left(1+\varepsilon\right)v/c}{1
-\left(1-\varepsilon\right)v/c}\right)^{c/2\kappa}
\frac{t+\left(1-\varepsilon^2\right)v\,x/c^2}{\sqrt{1+
2\,\varepsilon\,v/c-\left(1-\varepsilon^2\right)v^2/c^2}}}\\
\bar{x}={\displaystyle
\left(\frac{1+\left(1+\varepsilon\right)v/c}{1
-\left(1-\varepsilon\right)v/c}\right)^{c/2\kappa}
\frac{\left(1+2\,\varepsilon\,v/c\right)x+v\,t}{\sqrt{1+
2\,\varepsilon\,v/c-\left(1-\varepsilon^2\right)v^2/c^2}}}
\end{array}\right\}\;;
\label{L-trans}
\end{equation}
obviously, a generalisation of the standard Lorentz transformation.

In the cases $\varepsilon=\pm1$ ($a=0$) there is only one finite limit velocity, equal to $-1/b=-\varepsilon\,c/2=-\varepsilon\,C$, where $C:=c/2$ denotes the limit speed.  When $\varepsilon=-1$, $J=\,]-\infty,C\,[\,$, whereas for $\varepsilon=1$, $J=\,]-C,+\infty\,[\,$.  Equation (\ref{u}) gives
\begin{equation}
h(v)=\varepsilon\,C\ln\left(1+\varepsilon\,v/C\right)\;.
\label{h-case2}
\end{equation}
Of course, these can be considered as limit situations within the previous case, with either $c_-\to +\infty$ ($b<0$, $\varepsilon=-1$) or $c_+\to +\infty$ ($b>0$, $\varepsilon=1$).

\subsubsection{Case $\sigma=1$, $c=+\infty$ (or $\sigma=\varepsilon=0$)}
\label{subsubsec:3}

The limit velocities are both infinite, so $J=\mathbb{R}$ and $h(v)=v$.  The functions $s_1$ and $s_2$ are trivial: $s_1(v)=1$ and $s_2(v)=v$.  The transformation is
\begin{equation}
\left.\begin{array}{l}
\bar{t}={\displaystyle \ee^{v/\kappa}\,t}\\
\bar{x}={\displaystyle \ee^{v/\kappa}\left(x+v\,t\right)}
\end{array}\right\}\;,
\label{G-trans}
\end{equation}
which, apart from the factor $\ee^{v/\kappa}$, coincides with the Galilei transformation.  These results can all be recovered as the limit $c\to +\infty$ of those obtained in Sec.~\ref{subsubsec:1}.

\subsubsection{Case $\sigma=-1$, $c<+\infty$}
\label{subsubsec:6}

As already pointed out, there are no limit ``velocities''.  The functions $s_1$ and $s_2$ are
\begin{equation}
s_1(v)=\cos\left(h(v)/c\right)\;,\qquad\qquad\qquad
s_2(v)=c\,\sin\left(h(v)/c\right)\;,
\label{ss}
\end{equation}
and
\begin{equation}
\frac{v/c}{1+\varepsilon\,v/c}=\tan\left(h(v)/c\right)\;.
\label{h-case5}
\end{equation}
The transformation is better expressed in terms of the canonical parameter $V=h(v)$, since the use of $v$ would require some sign ambiguities in order to describe it completely: 
\begin{equation}
\left.\begin{array}{l}
c\,\bar{t}=\ee^{V/\kappa}\Big[\Big(\cos\left(V/c\right)-\varepsilon\,\sin\left(V/c\right)\Big)\,c\,t -\left(1+\varepsilon^2\right)\sin\left(V/c\right)\,x\Big]\\
\bar{x}=\ee^{V/\kappa}\Big[\sin\left(V/c\right)\,c\,t
+\Big(\cos\left(V/c\right)+\varepsilon\,\sin\left(V/c\right)\Big)\,x\Big]
\end{array}\right\}\;.
\label{E-trans}
\end{equation}
This generalises rotations in a Euclidean plane.  Pre-causality is violated for all choices of $\varepsilon$.   

\subsubsection{Case $\sigma=0$, $c<+\infty$, $\varepsilon\neq 0$}
\label{subsubsec:5}

The function $\varphi$ is
\begin{equation}
\varphi(v)=\left(1+b\,v/2\right)^2\;,
\label{phi-case3}
\end{equation}
and there is only one limit ``velocity'', $C:=-2/b$.  One finds
\begin{equation}
s_1(v)=1\;,\qquad s_2(v)=h(v)=\frac{v}{1-v/C}\;,
\label{h-case3}
\end{equation}
and the transformation is
\begin{equation}
\left.\begin{array}{l}
\bar{t}={\displaystyle \frac{\ee^{h(v)/\kappa}}{1-v/C}=\ee^{V/\kappa}\left(1+V/C\right)\,t}\\
\bar{x}={\displaystyle \ee^{h(v)/\kappa}\,\frac{\left(1-2\,v/C\right)\,x+v\,t}{1-v/C}=\ee^{V/\kappa}\Big(\left(1-V/c\right)\,x+V\,t\Big)}
\end{array}\right\}\;.
\label{schif-trans}
\end{equation}
Note that all these expressions can also be obtained by taking the limits $c\to +\infty$, $\varepsilon\to\pm\infty$, while keeping the ratio $\varepsilon/c$ finite, of those found in Sec.~\ref{subsubsec:6}.  Pre-causality is violated.

\subsection{Comments}
\label{subsec:interpretation}

Summarising, the only cases that correspond to physically acceptable kinematics are those considered in Secs.~\ref{subsubsec:1} and \ref{subsubsec:3}.  We now comment on the properties of the generalised composition law for velocities (Sec.~\ref{subsubsec:comp}), on the physical interpretation of the parameters $\varepsilon$ and $\kappa$ (Sec.~\ref{subsubsec:anisotropy}), and on the extension to three spatial dimensions (Sec.~\ref{subsubsec:3d}).

\subsubsection{Velocity composition law}
\label{subsubsec:comp}

The function $\Phi$ given in Eq.~(\ref{Phiab}) satisfies the following properties:
\begin{equation}
\Phi(u,0)=\Phi(0,u)=u\;,\qquad\forall u\in J\;;
\label{id}
\end{equation}
\begin{equation}
\forall u\in J,\;\exists\, u^\ast\in J\quad
\mbox{such that}\quad\Phi(u^\ast,u)=\Phi(u,u^\ast)=0\;;
\label{inverse}
\end{equation}
\begin{equation}
\Phi(\Phi(u,v),w)=\Phi(u,\Phi(v,w))\;,\qquad\forall u,v,w\in J\;;
\label{ass}
\end{equation}
\begin{equation}
\Phi(u,v)=\Phi(v,u)\;,\qquad\forall u,v\in J\;.
\label{comm}
\end{equation}
Hence, on writing $u\oplus v:=\Phi(u,v)$, $\forall u,v\in J$, Eq.\ (\ref{Phiab}) defines the composition law of an Abelian group $(J,\oplus)$, with neutral element $0$ and inverse $u^\ast$ of a generic element $u\in J$ defined by Eq.\ (\ref{inverse}).\footnote{In general, $u^\ast=-u/\left(1+b\,u\right)$, so $u^\ast=-u$ only when $b=0$.}  This can also be derived straightforwardly from simple kinematical arguments and the principle of relativity~\cite{sp}.  Remarkably, however, the associative and commutative properties (\ref{ass}) and (\ref{comm}) do not hold in general for the composition law of velocities along arbitrary directions in more than one spatial dimension~\cite{ungar-assoc}.

Since the function $\Phi$ is infinitely differentiable in both its arguments, $(J,\oplus)$ is a $C^\infty$ one-dimensional Lie group.  Any smooth, connected one-dimensional manifold is diffeomorphic either to $\mathbb{R}$, or to one of the real intervals $[0,1]$ and $[0,1[\,$, or to the circle $S^1$ (see, {\em e.g.\/}, Ref.~\cite{milnor}).  Hence, every one-dimensional Lie group is isomorphic either to $(\mathbb{R},+)$ or to $[0,1[\,$ with the addition modulo 1, according to its connectivity. This implies that, in our case, there exists a $C^\infty$ additive function defined on $J$, taking values either in $\mathbb{R}$ or in $[0,1[\,$, for which $0$ is a fixed point.\footnote{See Ref.~\cite{ll-p} for an alternative proof of this theorem, with applications to relativistic kinematics.}  Of course, this coincides with the function $h$ previously defined.  Since $h$ is one-to-one, $h(v)$ (called {\em rapidity\/} in the literature on special relativity) can be taken as an alternative mathematical representation for the physical notion of velocity, instead of the more common $v$.  This idea is supported by the observation that $h(v)$ also admits an operational definition~\cite{speed(s)}.

Suppose that a point has velocity $u$ with respect to the reference frame $\cal K$. Its velocity $\bar{u}$ with respect to $\overline{\cal K}$ is also given by the composition law (\ref{Phiab}), as one derives straightforwardly from the transformation equations (\ref{trans}), with $\xi(v)$ and $\eta(v)$ given by Eqs.~(\ref{xi}) and (\ref{eta}), respectively, with the obvious replacement $u\to v$.\footnote{Note that the function $A$ does not enter the velocity composition law.}  However, now only the velocity $v$ (of $\cal K$ with respect to $\overline{\cal K}$) is forced to belong to the interval $J$, because the moving point is not necessarily associated with some material particle.\footnote{It could correspond to a purely geometrical occurrence like, for example, the intersection between two moving straight lines, which can have an arbitrarily high speed even if the lines move rather slowly, provided the angle they form is small enough.}

The function $\varphi$ contains all the information needed to specify $\Phi$.  Its meaning can be found by expanding $\bar{u}$ to the first order in $v$:
\begin{equation}
\bar{u}=u+\varphi(u)\,v+{\cal O}(v^2)\;.
\label{expu}
\end{equation}
This is the composition law between an {\em arbitrary\/} velocity $u$ and a velocity $v$ with {\em small\/} magnitude.  Since Eq.\ (\ref{id}) implies $\varphi(0)=1$, at very small speeds one always recovers Galilean kinematics.  However, unless $\varphi(u)\equiv 1$, deviations from the Galilean composition law can always be detected if one measures the speed of a fast-moving object, even for small values of the relative velocity $v$ between frames.

Another interesting property of $\varphi$ is that if some finite velocity, say $C$, is invariant, then $\varphi(C)=0$. This follows immediately by applying Eq.~(\ref{f(u)}) to the condition
\begin{equation}
\Phi(C,v)=C\;,\qquad\forall v\in J\;,
\label{fixed}
\end{equation}
which expresses the invariance of $C$.  Hence, the limit velocities for reference frames coincide with the invariant velocities.\footnote{Note that the possibility for the existence of invariant speeds has been derived as a kinematical possibility only from the postulates of relativity, homogeneity, and pre-causality.  This approach to relativistic kinematics was pioneered by von Ignatowsky in 1910~\cite{ignat}, and was later rediscovered many times in different ways~\cite{lalan,ll-p,rediscoverer,jammer}. See also~\cite{gorini} for a rigorous treatment, and~\cite{textbooks,brown} for clear presentations at a textbook level.}

\subsubsection{Conventional and real anisotropy}
\label{subsubsec:anisotropy}

For non-vanishing values of the quantity $\varepsilon\in\,]-1,1[\,$, the invariant speeds $c_\pm$ along the two orientations of one-dimensional space differ from each other~\cite{one-way}.  On the other hand, the average (two-ways) speed of a signal travelling at the invariant velocity along a round trip turns out to be equal to the $\varepsilon$-independent parameter $c$.  For $\varepsilon\to\pm 1$, one of the two invariant speeds tends to infinity, while the Galilean composition law is recovered in the limit $c\to +\infty$.  A further requirement of spatial isotropy (or better, its one-dimensional counterpart --- the physical equivalence of the two orientations in the one-dimensional space) enforces, not surprisingly, the equality between $c_+$ and $c_-$~\cite{lalan}.  Hence, anisotropy seems to be related to the possibility of having $c_+\neq c_-$.  However, whether the one-way speed of light is a physically meaningful quantity or merely a conventional one, is the matter of a long-standing debate~\cite{one-way,synchro}.  

This issue is inextricably linked to another one, concerning the conventionality of clock synchronisation~\cite{jammer,brown,synchro}.  Any measurement of the one-way speed of light requires, in order to be performed, a prior synchronisation of distant clocks.  Vice versa, one may argue that any synchronisation procedure is equivalent to a stipulation about the value of the one-way speed of light.\footnote{These statements are less tautological than it may seem at first, as one can synchronise without using light.}  This creates a circularity which does not allow any escape.  Indeed, as it is evident from the first of equations (\ref{L-trans}), $\varepsilon$ also encodes the convention adopted for synchronising clocks.  The value $\varepsilon=0$ corresponds to the usual (Einstein's) procedure, while other values account for Reichenbach's generalised synchronisations~\cite{jammer,brown,synchro}.  Thus, for synchronisation procedures different from Einstein's, $c_+\neq c_-$, whereas the value of $c$ does not depend on the synchronisation procedure adopted. 

At a first sight, this circumstance appears puzzling.  Isotropy, being related with the equality of the one-way invariant speeds $c_\pm$, can be regarded as a consequence of Einstein's  synchronisation procedure, and it has been argued convincingly that the choice of such a procedure is merely conventional, in spite of some claims to the contrary.\footnote{See~\cite{brown,synchro,torretti-phil} for the debate.}  On the other hand, it seems reasonable to believe that isotropy (or anisotropy) is a physical property, which cannot be implemented or altered just by a stipulation.  

The resolution of this conundrum is that $\varepsilon=0$ is only a necessary, but not sufficient, condition for isotropy.  The transformation~(\ref{L-trans}) contains also the parameter $\kappa$, and if $|\kappa|<+\infty$ one can experimentally distinguish between the two spatial orientations even if $\varepsilon=0$.  For example, if one considers two clocks moving at the same speed along opposite directions, these clocks will not delay by the same amount, if $|\kappa|<+\infty$. Hence, anisotropy can manifest itself through physical effects, independent of the choice of synchronisation.  One should thus distinguish between a {\em real\/}, physical anisotropy, measured by the parameter $\kappa$, and a purely {\em conventional\/} one --- that can be introduced or gauged away simply by a stipulation --- measured by $\varepsilon$.  From this point of view, the fact that one may not use Einstein's synchronisation is a trivial one, and has no physical content.  On the contrary, the fact that one {\em can\/} adopt it in any inertial frame is physically non-trivial.

An analogy to this situation is provided by the principle of inertia.  The statement that, in an inertial frame, a force-free particle moves along a straight line at a constant speed contains two elements, very different in nature.  One, that the motion takes place along a straight line, is a physically testable prediction, since the notion of a straight line is well defined in the Euclidean geometry that one presupposes valid when formulating the principle.  The other, that such motion is uniform, is a matter of convention.  Of course, one could choose the ``time'' variable in such a way that the motion is {\em not\/} uniform (the analog of choosing a synchronisation different from Einstein's), but this generalisation will not lead to new physical phenomena --- only to a horrendous complication in the formulation of the laws of mechanics.  Again, the relevant fact is not that one can make an absurdly complicated choice of time, but rather that one can make a choice that simplifies life.  Since a change in the time variable does not entail new phenomena, we are confronted with a mere gauge, and the wisest choice is to use the simplest possible gauge.  As Misner, Thorne and Wheeler concisely and effectively wrote: ``Time is defined so that motion looks simple''~\cite{mtw}.  We could paraphrase them saying: ``Clocks are synchronised so that physics looks simple.''

Intriguingly, there is a possibility that the value of $\kappa$ also reflects a mere convention.  The generalised Lorentz transformation (\ref{L-trans}) with $\varepsilon=0$ is a standard Lorentz transformation accompanied by a global dilatation by a $\kappa$-dependent factor, dilatation which can be attributed to a rescaling of the time and distance units in the reference frame $\overline{\cal K}$.  Such a rescaling could be an effect of anisotropy on clocks and rulers, but could also be induced artificially, by a suitable choice of units in different frames.  In any case, it can be compensated through a rescaling of units in the reference frame $\overline{\cal K}$.\footnote{This operation does not destroy the group structure, and is therefore compatible with the principle of relativity.  On the contrary, eliminating the Lorentz factor by a rescaling of units would not preserve the group structure, hence would imply a violation of the principle.}  

There is a way out of this difficulty, though.  The anisotropic scale factor in Eq.~(\ref{L-trans}) is very special, and requires a carefully tailored choice of units in $\overline{\cal K}$ in order to be produced artificially.    In common experimental practice, there is a well-defined (although not explicitly stated) procedure for building units in a frame $\overline{\cal K}$, say.  They are either constructed directly in $\overline{\cal K}$ following some standard instructions; or they are boosted from another frame $\cal K$, where they have been built using the standard instructions.  (That these two operations produce the same units in $\overline{\cal K}$, is the so-called principle of the boostability of units~\cite{brown}.)  Thus, although the choice of units in different frames is, in principle, free, in practice it is always made in the same way --- a way, moreover, which does not seem capable of introducing an anisotropy.  Hence, any detection of the anisotropy allowed by Eq.~(\ref{L-trans}) will be regarded as a real physical effect, rather than as an artifact of conventions.\footnote{The situation is different for the transformation (\ref{E-trans}).  When $\varepsilon=0$ this is a rotation, anticlockwise by an angle $\theta:=V/c$, of orthogonal axes in a Euclidean plane, accompanied by a global dilatation by the factor ${\rm e}^{\theta c/\kappa}$.  A true anisotropy cannot produce effects for $\theta=2\pi$, so the factor ${\rm e}^{\theta c/\kappa}$ can only be due to a devious choice of units in the frame $\overline{\cal K}$.}  Such an effect could be gauged away only at the price of changing everyone's measurement habits.

\subsubsection{Three-dimensional case}
\label{subsubsec:3d}

Although our treatment was restricted to the case of one spatial dimension, it is easy to extend it to a three-dimensional space.  We do not provide a systematic generalisation here, but nevertheless we wish to present a physical argument that shows what one could expect.  (See~\cite{bogoslovsky,bogo} for more details.)  For simplicity, we choose Einstein's  synchronisation ($\varepsilon=0$).  Consider a light-beam clock, of the type that is  commonly used in pedagogical treatments of special relativity (see, {\em e.g.\/}, Ref.~\cite{merminbook}).  Basically, such a clock is made of two mirrors, one along the $x$-axis, the other displaced from it along the perpendicular direction.  Let us denote by $y$ and $\bar{y}$ the distance between the mirrors in the reference frames $\cal K$ and $\overline{\cal K}$, respectively.  The time taken by light to make a complete two-way trip between the mirrors is, in $\cal K$, simply $\Delta t=2\,y/c$.  The corresponding time according to $\overline{\cal K}$ is then given by the first equation in (\ref{L-trans}) with $\varepsilon=0$:
\begin{equation}
\Delta\bar{t}=\left(\frac{1+v/c}{1-v/c}\right)^{c/2\kappa}\frac{2\,y/c}{\sqrt{1-v^2/c^2}}\;.
\label{Deltat}
\end{equation}
But $\Delta\bar{t}$ can also be found directly from the equation
\begin{equation}
\Delta\bar{t}=\frac{2}{c}\,\sqrt{\left(v\,\Delta\bar{t}/2\right)^2+\bar{y}^2}\;,
\label{Deltatbar}
\end{equation}
which gives
\begin{equation}
\Delta\bar{t}=\frac{2\,\bar{y}/c}{\sqrt{1-v^2/c^2}}\;.
\label{Deltatbar!}
\end{equation}
Combining Eq.~(\ref{Deltat}) with $\Delta t=2\,y/c$ and Eq.~(\ref{Deltatbar!}), we find 
\begin{equation}
\bar{y}=\left(\frac{1+v/c}{1-v/c}\right)^{c/2\kappa} y\;,
\label{ybar}
\end{equation}
instead of the usual $\bar{y}=y$.  Thus, anisotropy introduces also a transformation for distances along transverse directions.

\section{Foundations for dynamics}
\label{sec:general}
\setcounter{equation}{0}

We now present a method for finding the basic dynamical quantities (Lagrangian, energy and momentum) for a free particle, that are compatible with the kinematics presented in Sec.~\ref{sec:kin}.  

\subsection{Lagrangian}
\label{subsec:lagrangian}

The Lagrangian $L$ for a free particle in an inertial frame can be determined, in Newtonian dynamics, by the following argument~\cite{ll}. Space and time homogeneity requires that $L$ do not depend on the particle position and on time, so it must be only a function of velocity. By the relativity principle, we have that in the inertial frames $\cal K$ and $\overline{\cal K}$ this function must be the same, so we shall write $L(u)$ and $L(\bar{u})$, respectively, where $\bar{u}=u+v$. However, these two Lagrangians must lead to the same equation of motion, so they can differ by a total derivative with respect to time of a function $f$ of the particle coordinate and of time.  Hence 
\begin{equation}
L(u+v)=L(u)+\frac{\partial f(x,t;v)}{\partial x}\,u+\frac{\partial
f(x,t;v)}{\partial t}\;.
\label{LNewt}
\end{equation}
Since the left-hand side does not depend on $x$ and $t$, the same must happen for the right-hand side, so the quantities $\partial f/\partial x$ and $\partial f/\partial t$ can actually be functions of $v$ only.  Denoting them as $\alpha(v)$ and $\beta(v)$, we have therefore 
\begin{equation}
L(u+v)=L(u)+\alpha(v)\,u+\beta(v)\;.
\label{LNewt'}
\end{equation}
Taking the derivative of Eq.~(\ref{LNewt'}) with respect to $v$, and evaluating the result for $v=0$, we get
\begin{equation}
L'(u)=\alpha'(0)\,u+\beta'(0)\;.
\label{LN'}
\end{equation}
Integrating, one finds that $L(u)$ is a polynomial of second degree in $u$.  Discarding the part linear in $u$, which does not contribute to the equation of motion, one ends up with the usual result: $L(u)$ is proportional to $u^2$.  Note that this implies isotropy of space, although no such hypothesis has been explicitly used.

It is very easy to see that one cannot run this argument in a straightforward manner to get the Lagrangian corresponding to a generic composition law (\ref{comp-gen}).  Indeed, even for the composition law of special relativity one would not recover the standard expression for the Lagrangian.  The reason lies in the fact that Eq.~(\ref{LNewt}) is appropriate only for a theory in which the time coordinates $t$ and $\bar{t}$ in $\cal K$ and $\overline{\cal K}$, respectively, coincide.  As we saw in Sec.~\ref{sec:kin}, this happens only in isotropic Galilean kinematics.

This difficulty can be overcome by working in the extended configuration space of the particle, where position and time are both treated as Lagrangian coordinates, evolving in terms of a parameter $\theta$~\cite{lanczos}.  The condition that the Lagrangian, expressed in the inertial frames $\cal K$ and $\overline{\cal K}$, lead to the same equation of motion is then
\begin{equation}
\frac{\d\bar{t}}{\d\theta}L\left(\frac{\d\bar{x}}{\d\theta}
\left/\right.\frac{\d\bar{t}}{\d\theta}\right) =\frac{\d
t}{\d\theta}L\left(\frac{\d x}{\d\theta}\left/\right.\frac{\d
t}{\d\theta}\right)+\alpha(v)\,\frac{\d
x}{\d\theta}+\beta(v)\,\frac{\d
t}{\d\theta}\;,
\label{ext}
\end{equation}
where $x$ and $\bar{x}$ denote the particle position in $\cal K$ and $\overline{\cal K}$, respectively.  Equation (\ref{ext}) can be rewritten as
\begin{equation}
g(u,v)L(\Phi(u,v))=L(u)+\alpha(v)\,u+\beta(v)\;,
\label{ext'}
\end{equation}
where we have defined the function
\begin{equation}
g(u,v):=\frac{\d\bar{t}}{\d\theta}\left/\right.\frac{\d t}{\d\theta}\;,
\label{g}
\end{equation}
such that $g(u,0)=1$.

We can now proceed as in the Galilean case, by taking the derivative of Eq.~(\ref{ext'}) and evaluating it for $v=0$.  The result is the following first-order ordinary linear differential equation for the function $L(u)$:
\begin{equation}
\varphi(u)\,L'(u)+\rho(u)\,L(u)=\alpha'(0)\,u+\beta'(0)\;,
\label{eqforL}
\end{equation}
where
\begin{equation}
\rho(u):=\left.\frac{\partial g(u,v)}{\partial v}\right|_{v=0}\;.
\label{rho}
\end{equation}
Using Eqs.~(\ref{trans-matrix}), (\ref{lambda}), (\ref{A}) and (\ref{xi}) into Eq.~(\ref{g}), we now find immediately
\begin{equation}
g(u,v)=\left(1+a\,u\,v\right)A(v)\;.
\label{gtrue}
\end{equation}
Hence, remembering that $A(0)=1$,
\begin{equation}
\rho(u)=a\,u+A'(0)=a\,u+1/\kappa-b/2\;,
\label{g'0}
\end{equation}
and the differential equation (\ref{eqforL}) becomes then
\begin{equation}
\left(1+b\,u-a\,u^2\right)L'(u)+\left(a\,u+1/\kappa-b/2\right)L(u)
=\alpha'(0)\,u+\beta'(0)\;.
\label{eqforL'}
\end{equation}

It would now be straightforward to find the Lagrangian corresponding to the various possible kinematics classified in Sec.~\ref{sec:kin}.  However, we find it interesting to show that the {\em same\/} functions $\varphi$ and $\rho$ can be obtained {\em without\/} any explicit reference to the derivation in Sec.~\ref{subsec:trans}, simply requiring that the dynamics based on the Lagrangian $L$ be compatible with the existence of elastic collisions between asymptotically free particles.

\subsection{Momentum and energy}
\label{subsec:momentum}

Given the Lagrangian $L(u)$, the particle momentum and energy\footnote{Hereafter, whenever we refer to ``energy'' we mean the sum of kinetic energy and a possible rest energy.} are easily found by the standard relations:
\begin{equation}
p(u)=L'(u)\;;
\label{p=dL/du}
\end{equation}
\begin{equation}
E(u)=u\,p(u)-L(u)\;.
\label{E=}
\end{equation}

Noteworthy, expressions for $p(u)$ and $E(u)$ can also be derived following the procedure in Ref.~\cite{sp}, which relies directly on the existence of elastic collisions and on the relativity principle,\footnote{Basically, a straightforward generalisation of an argument originally due to Huygens~\cite{barbour}.  See also~\cite{davidon,levy} for similar developments.} instead of starting from a Lagrangian.  Let $E(u)$ be the energy of a particle with velocity $u$ in an inertial frame $\cal K$.  In an inertial frame, $E(u)$ is conserved for a free particle, because $u$ is constant, by the principle of inertia. Then, the total energy --- defined as the sum of the energies for the individual particles --- is conserved also for a system of noninteracting particles.

Let us assume that there are spatially localised interactions between particles which do not change the total energy.  Then, during such an interaction between two particles with initial velocities $u_1^{({\rm i})}$, $u_2^{({\rm i})}$, and final velocities $u_1^{({\rm f})}$, $u_2^{({\rm f})}$:
\begin{equation}
E_1(u_1^{({\rm i})})+E_2(u_2^{({\rm i})})=E_1(u_1^{({\rm f})})
+E_2(u_2^{({\rm f})})\;.
\label{T+T}
\end{equation}
(Of course, the energy may depend on some invariant parameter characterising the particle, in addition to its velocity.  For example, in Newtonian dynamics it depends on the particle mass. We keep track of this dependence through the indices 1 and 2 on $E$.)  With respect to another inertial frame $\overline{\cal K}$,
\begin{equation}
E_1(\bar{u}_1^{({\rm i})})+E_2(\bar{u}_2^{({\rm i})})
=E_1(\bar{u}_1^{({\rm f})})+E_2(\bar{u}_2^{({\rm f})})\;,
\label{T'+T'}
\end{equation}
where the various $\bar{u}^{({\rm i})}$ and $\bar{u}^{({\rm f})}$ are given by the composition law for velocities (\ref{comp-gen}), and we have used the same functions $E_1$ and $E_2$ in both reference frames because of the relativity principle.

We now expand the generic functions in Eq.~(\ref{T'+T'}) around $v=0$, and use Eq.~(\ref{expu}) to get
\begin{equation}
E(\bar{u})=E(u)+E'(u)\,\varphi(u)\,v+{\cal O}(v^2)\;.
\label{exp}
\end{equation}
Doing this for each term in Eq.~(\ref{T'+T'}) and using Eq.~(\ref{T+T}), then dividing by $v$ and taking the limit for $v\to 0$, we obtain
\begin{equation}
E'_1(u_1^{({\rm i})})\,\varphi(u_1^{({\rm i})})+E'_2(u_2^{({\rm
i})})\,\varphi(u_2^{({\rm i})})=
E'_1(u_1^{({\rm f})})\,\varphi(u_1^{({\rm f})})
+E'_2(u_2^{({\rm f})})\,\varphi(u_2^{({\rm f})})\;.
\label{consmom}
\end{equation}
Hence, there is another additive quantity which is conserved, in addition to energy.  For a single particle, the most general expression for such a quantity is
\begin{equation}
p(u)=\lambda\,\varphi(u)\,E'(u)+\mu\,E(u)+\nu\;,
\label{momentum}
\end{equation}
where $\lambda$, $\mu$, and $\nu$ are quantities independent of $u$.  The function $p(u)$ given by Eq.~(\ref{momentum}) with $\lambda=1$, $\mu=\nu=0$ coincides with linear momentum\footnote{Note that with this identification, linear momentum turns out to be (correctly) a one-form rather than a vector~\cite{sp-3d}.} both in Newtonian and Einstein mechanics~\cite{sp}, and we shall retain such an interpretation for the more general expression above.\footnote{No new conservation laws can arise at higher orders in $v$, as argued by L\'evy-Leblond~\cite{levy}.  See also Ref.~\cite{sp} for an explicit proof when $E$ is only a function of $u^2$ (which, however, is not the case in the presence of anisotropy).}

\subsection{Compatibility}
\label{subsec:compat}

If one assumes that $p$ and $E$ can be derived from a Lagrangian according to Eqs.~(\ref{p=dL/du}) and (\ref{E=}), Eq.~(\ref{momentum}) can be converted into a second-order differential equation for $L$:
\begin{equation}
\lambda\,u\,\varphi(u)\,L''(u)+\left(\mu\,u-1\right)\,L'(u)-
\mu\,L(u)+\nu=0\;.
\label{2L}
\end{equation}
In order for the treatments in Secs.~\ref{subsec:lagrangian} and~\ref{subsec:momentum} to be mutually compatible, Eqs.~(\ref{eqforL}) and (\ref{2L}) must have the same content. To compare them, let us first rewrite Eq.~(\ref{eqforL}) as a second-order differential equation.  By taking a derivative with respect to $u$, then multiplying by $u$, we obtain
\begin{equation}
u\,\varphi(u)\,L''(u)+u\left(\varphi'(u)+\rho(u)\right)\,L'(u)
+u\,\rho'(u)\,L(u)-\alpha'(0)\,u=0\;.
\label{derL}
\end{equation}
We can now replace the last term in Eq.~(\ref{derL}), $\alpha'(0)\,u$, by using again Eq.~(\ref{eqforL}).  The final equation is
\begin{equation}
u\,\varphi(u)\,L''(u)+\left(u\,\varphi'(u)-\varphi(u)
+u\,\rho(u)\right)\,L'(u)+\left(u\,\rho'(u)
-\rho(u)\right)\,L(u)+\beta'(0)=0\;.
\label{3L}
\end{equation}

Equations~(\ref{2L}) and (\ref{3L}) coincide if the following relations hold:
\begin{equation}
\beta'(0)=\nu/\lambda\;;
\label{1}
\end{equation}
\begin{equation}
u\,\rho'(u)-\rho(u)=-\mu/\lambda\;;
\label{2}
\end{equation}
\begin{equation}
u\,\varphi'(u)-\varphi(u)+u\,\rho(u)
=\mu\,u/\lambda-1/\lambda\;.
\label{3}
\end{equation}
The most general form of the functions $\rho$ and $\varphi$ allowed by Eqs.~(\ref{2}) and (\ref{3}) are
\begin{equation}
\rho(u)=a\,u+\mu/\lambda
\label{genrho}
\end{equation}
and
\begin{equation}
\varphi(u)=-a\,u^2+b\,u+1/\lambda\;,
\label{genphi}
\end{equation}
with $a$ and $b$ arbitrary constants.\footnote{The constants $a$ and $b$ have, {\em a priori\/}, nothing to do with those introduced in Sec.~\ref{subsec:trans}, but we shall soon discover that they actually coincide. This justifies using the same letters in the notation.}  Moreover, Eq.~(\ref{id}) implies $\varphi(0)=1$, so $\lambda=1$.  Hence, the treatments in Secs.~\ref{subsec:lagrangian} and~\ref{subsec:momentum} are compatible only if:
\begin{equation}
\rho(u)=a\,u+\mu\;;
\label{rho1}
\end{equation}
\begin{equation}
\varphi(u)=-a\,u^2+b\,u+1\;;
\label{phiab}
\end{equation}
\begin{equation}
\nu=\beta'(0)\;.
\label{nu}
\end{equation}
Remarkably, the structure of $\rho$ and $\varphi$ emerges by the requirement that the treatments in Secs.~\ref{subsec:lagrangian} and \ref{subsec:momentum} be compatible, with no independent considerations about kinematics.  Note that the expression for $\varphi(u)$ coincides with the one given by Eq.~(\ref{phiab'}), obtained on purely kinematical grounds.  Comparing now Eq.~(\ref{rho1}) with Eq.~(\ref{g'0}), we are led to the identification
\begin{equation}
\mu=A'(0)=1/\kappa-b/2\;.
\label{muu}
\end{equation}
%

\subsection{Mass, rest energy, rest momentum}
\label{subsec:mass}

Evaluating Eq.~(\ref{eqforL'}) and its first derivative at $u=0$, we can express the quantities $\alpha'(0)$ and $\beta'(0)$ in terms of $a$, $b$, $\mu$, and of the three parameters $m:=L''(0)$, $p_0:=p(0)=L'(0)$, and $E_0:=E(0)=-L(0)$, that represent the particle's mass and possible rest momentum and energy, respectively.\footnote{One could distinguish between the momentum $p$ and a ``kinetic momentum'' $p-p_0$, just as one usually distinguishes between the energy $E$ and the ``kinetic energy'' $E-E_0$.}  The results are
\begin{equation}
\alpha'(0)=m+(b+\mu)\,p_0-a\,E_0
\label{alpha'}
\end{equation}
and
\begin{equation}
\beta'(0)=p_0-\mu\,E_0\;,
\label{beta'}
\end{equation}
so the differential equation for $L$ is, finally:
\begin{equation}
\left(1+b\,u-a\,u^2\right)L'(u)+\left(a\,u+\mu\right)L(u)
=\left(m+(b+\mu)\,p_0-a\,E_0\right)\,u+p_0-\mu\,E_0\;.
\label{eqforLfinal}
\end{equation}
This is the equation we shall use as a starting point in the next section.  

It is important to note that the parameter $m$ coincides with the Newtonian mass, thus justifying our setting $m:=L''(0)$.\footnote{It might have been logically possible that $L''(0)$ were equal to the Newtonian mass multiplied by a function of $c$ and $\kappa$ that reduces to 1 when both these parameters tend to infinity.}  This can be seen by expanding the Lagrangian and keeping only the leading order terms in $u/c$ and $u/\kappa$.  The result is
\begin{equation}
L(u)=\frac{1}{2}\,mu^2+p_0u-E_0+{\cal O}(u^3)\;,
\label{Lexp}
\end{equation}
which to the first significant order in $u$ describes indeed a Newtonian particle with mass $m$.

Instead of solving Eq.~(\ref{eqforLfinal}), one could proceed as follows.  Differentiating Eq.~(\ref{E=}), and using Eq.~(\ref{p=dL/du}), one obtains the well-known relation
\begin{equation}
\d E(u)=u\,\d p(u)\;,
\label{dT}
\end{equation}
expressing the fact that the change in the energy of a particle equals the work done on it --- an elementary property that holds not only in Newtonian and Einstein mechanics, but in any dynamics that admits a Lagrangian formulation.  Combining Eq.~(\ref{momentum}) with $\lambda=1$ and Eq.~(\ref{dT}), one obtains a single differential equation, which can be solved to find the expression for $E(u)$.  Inserting the latter into Eq.~(\ref{momentum}), one finds also the expression for $p(u)$. 

Finally, let us notice that momentum and energy must be proportional to the particle mass $m$.  This follows from the requirement that momentum, energy, and mass are all additive quantities.  Denoting by $p(m,u)$ and $E(m,u)$ the momentum and energy of a particle with mass $m$ and velocity $u$, we have thus, for a system of two particles with masses $m_1$ and $m_2$ that move with the same velocity $u$,
\begin{equation}
p(m_1,u)+p(m_2,u)=p(m_1+m_2,u)\;,
\label{sump}
\end{equation}
and
\begin{equation}
E(m_1,u)+E(m_2,u)=E(m_1+m_2,u)\;,
\label{sumE}
\end{equation}
from which the said proportionality easily follows.  As a byproduct, the rest momentum $p_0$ and the rest energy $E_0$ turn out to be also proportional to $m$.

\section{Anisotropic relativistic dynamics}
\label{sec:examples}
\setcounter{equation}{0}

We now write the explicit expressions for the basic quantities that appear in an anisotropic relativistic dynamics.  Although the considerations in Sec.~\ref{subsubsec:anisotropy}  strongly suggest to set $\varepsilon$ to zero, and experimental evidence suggests $\kappa>c$, in the following we shall consider all the values of the parameters that correspond to possible kinematics, in order to remain as general as possible.  The differential equation (\ref{eqforLfinal}) has two different solutions when $|\kappa|\neq c$ and when $|\kappa|=c$, so we must consider these two cases separately.

\subsection{Case $|\kappa|\neq c$}
\label{subsec:kneqc}

The solution of Eq.~(\ref{eqforLfinal}) that satisfies the condition $L''(0)=m$ is
\begin{equation}
L(u)=-\frac{m}{1/c^2-1/\kappa^2}\,{\rm e}^{-h(u)/\kappa}\,\varphi(u)^{1/2}+\left(p_0-m\,\frac{1/\kappa-\varepsilon/c}{1/c^2
-1/\kappa^2}\right)u
-\left(E_0-\frac{m}{1/c^2-1/\kappa^2}\right)\;.
\label{Lknotc}
\end{equation}
The expressions for momentum and energy can then be obtained by Eqs.~(\ref{p=dL/du}) and (\ref{E=}):
\begin{equation}
p(u)=\frac{m}{1/c^2-1/\kappa^2}\,{\rm e}^{-h(u)/\kappa}\,\varphi(u)^{-1/2}\left(\frac{1-\varepsilon^2}{c^2}\,u+\frac{1}{\kappa}-\frac{\varepsilon}{c}\right)+\left(p_0-m\,\frac{1/\kappa-\varepsilon/c}{1/c^2-1/\kappa^2}\right)\;;
\label{pknotc}
\end{equation}
\begin{equation}
E(u)=\frac{m}{1/c^2-1/\kappa^2}\,{\rm e}^{-h(u)/\kappa}\,\varphi(u)^{-1/2}\left(1+\left(\frac{1}{\kappa}+\frac{\varepsilon}{c}\right) u\right)+\left(E_0-\frac{m}{1/c^2-1/\kappa^2}\right)\;.
\label{Eknotc}
\end{equation}
%
\subsubsection{Anisotropic Einstein's dynamics}
\label{subsec:1}

\begin{figure}[htbp]
\vbox{ \hfil
\scalebox{0.32}{{\includegraphics{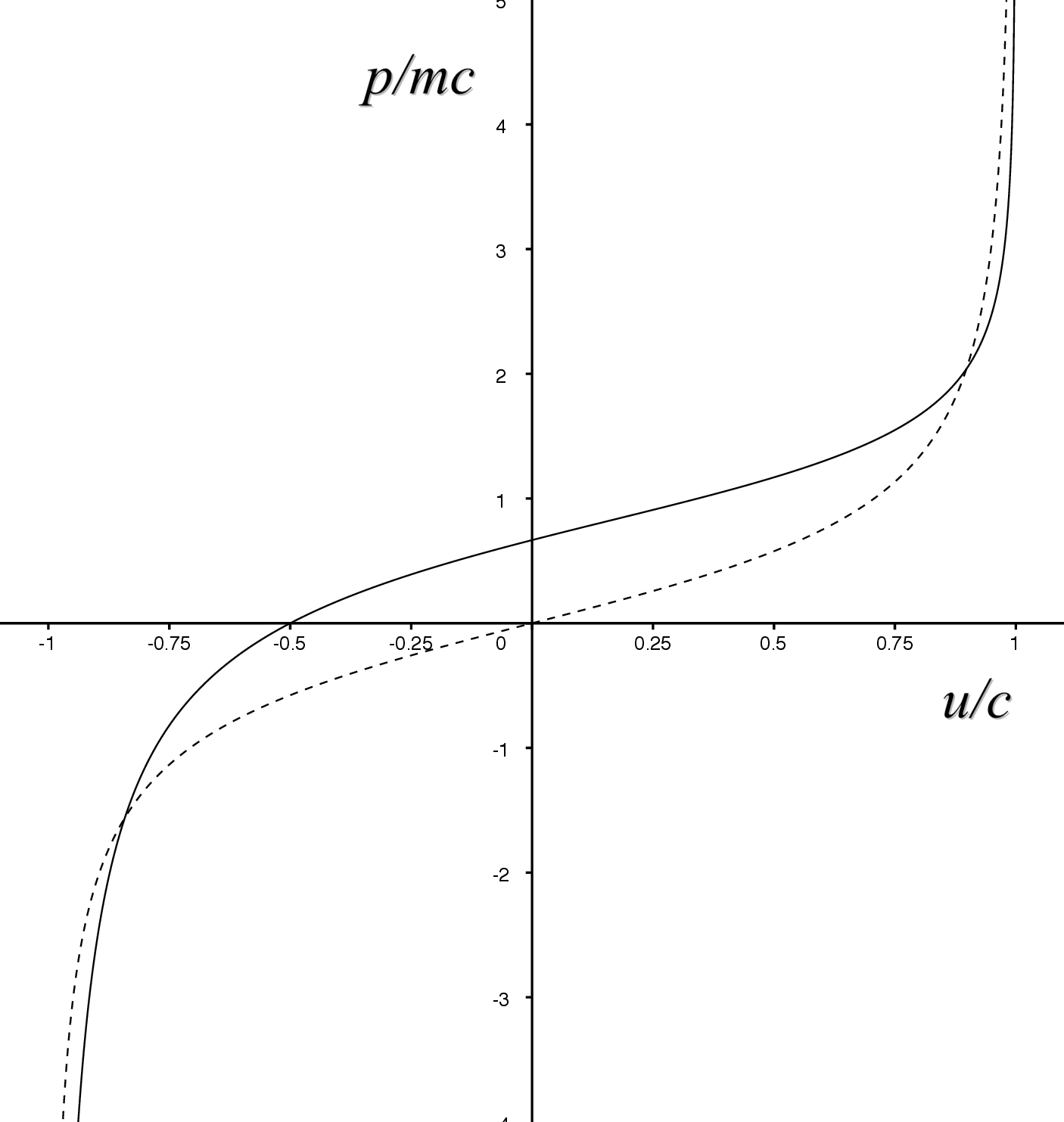}},\hskip2cm{\includegraphics{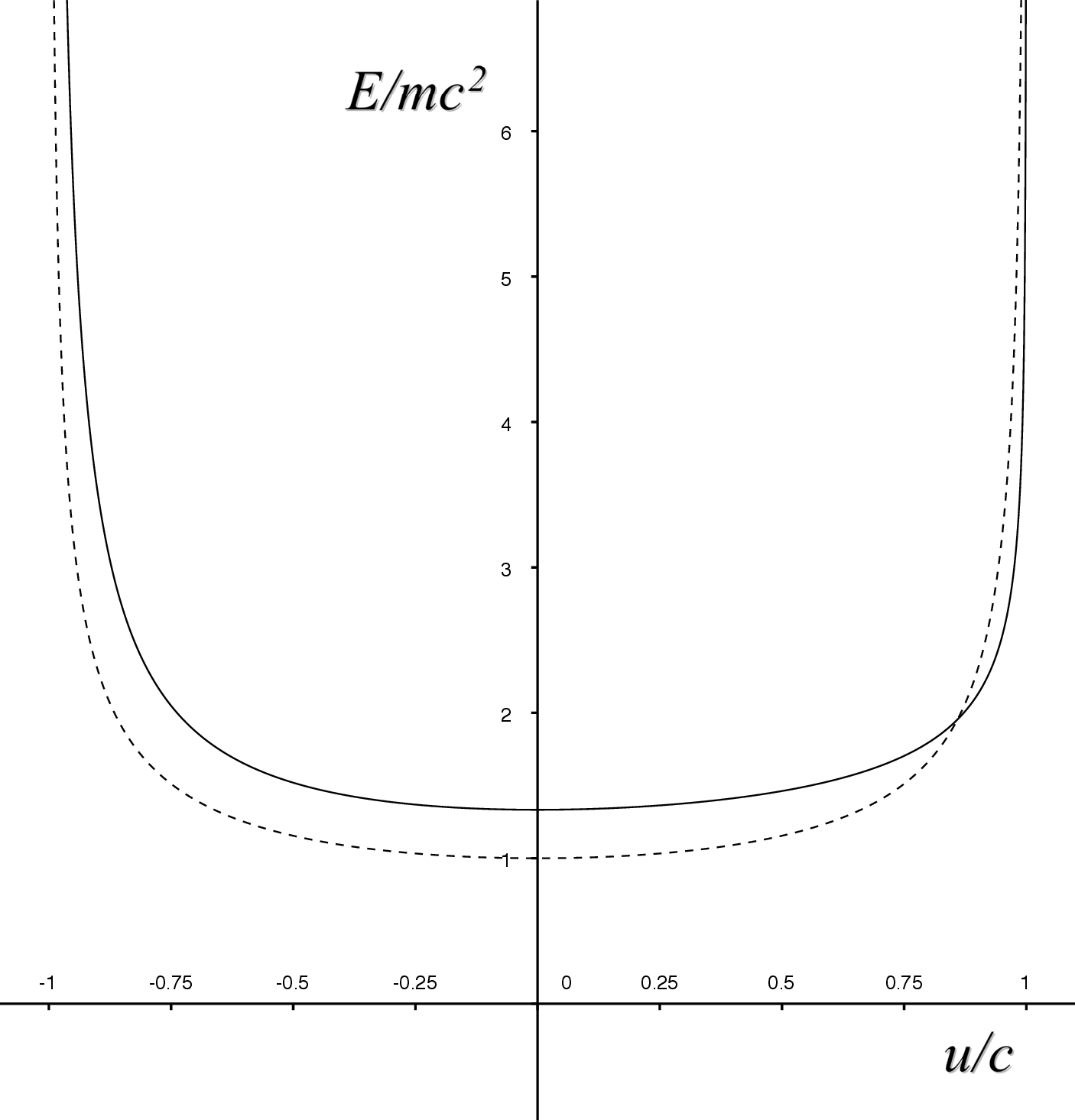}}}
\hfil }
\vskip.25cm
\caption{\small Comparison between the expressions given by Eqs.~(\ref{pEinst}) and (\ref{EEinst}) for $\kappa=2\,c$ (solid lines) and $|\kappa|=+\infty$ (dashed lines).  In both cases, the conventional anisotropy parameter $\varepsilon$ has been set equal to zero, and the values of the quantities at rest have been chosen according to Eqs.~(\ref{p0}) and (\ref{Emc2}). The plot on the left represents momentum, the one on the right represents energy.}
\label{fig1}
\end{figure}

Replacing Eq.~(\ref{h-case1}) into Eqs.~(\ref{Lknotc})--(\ref{Eknotc}) we find:
\begin{eqnarray}
L(u)=&-&\frac{m}{1/c^2-1/\kappa^2}
\left(\frac{1-\left(1-\varepsilon\right)u/c}{1
+\left(1+\varepsilon\right)u/c}\right)^{c/2\kappa}
\left(1-\left(1-\varepsilon\right)u/c\right)^{1/2}
\left(1+\left(1+\varepsilon\right)u/c\right)^{1/2}\nonumber\\
&+&\left(p_0-m\,\frac{1/\kappa-\varepsilon/c}{1/c^2
-1/\kappa^2}\right)u
-\left(E_0-\frac{m}{1/c^2-1/\kappa^2}\right)\;;
\label{LEinst}
\end{eqnarray}
\begin{eqnarray}
p(u)=&&\frac{m}{1/c^2-1/\kappa^2}
\left(\frac{1-\left(1-\varepsilon\right)u/c}{1
+\left(1+\varepsilon\right)u/c}\right)^{c/2\kappa}
\frac{\left(1-\varepsilon^2\right)u/c^2
-\varepsilon/c+1/\kappa}{\left(1-\left(1-\varepsilon\right)u/c\right)^{1/2}
\left(1+\left(1+\varepsilon\right)u/c\right)^{1/2}}\nonumber\\
&+&p_0-m\,\frac{1/\kappa-\varepsilon/c}{1/c^2-1/\kappa^2}\;;
\label{pEinst}
\end{eqnarray}
\begin{eqnarray}
E(u)=&&\frac{m}{1/c^2-1/\kappa^2}
\left(\frac{1-\left(1-\varepsilon\right)u/c}{1
+\left(1+\varepsilon\right)u/c}\right)^{c/2\kappa}
\frac{1+\left(1/\kappa+\varepsilon/c\right)u}{\left(1
-\left(1-\varepsilon\right)u/c\right)^{1/2}
\left(1+\left(1+\varepsilon\right)u/c\right)^{1/2}}\nonumber\\
&+&E_0-\frac{m}{1/c^2-1/\kappa^2}\;.
\label{EEinst}
\end{eqnarray}
These equations generalise the basic expressions of Einstein's dynamics (corresponding to $\varepsilon=0$, $|\kappa|=+\infty$) to the anisotropic case.  The cases in which anisotropy is merely due to a convention ($\varepsilon\neq 0$) and in which it is an intrinsic physical property ($|\kappa|< +\infty$) are both covered.  Figure~\ref{fig1} shows a comparison between these expressions and those valid  assuming isotropy.  
Note that, even in the presence of anisotropy, the minimum value of $E$ is still attained for $u=0$.

Remarkably, Eqs.~(\ref{pEinst}) and (\ref{EEinst}) suggest both the possibility for a non-vanishing rest momentum
\begin{equation}
p_0=m\,\frac{1/\kappa-\varepsilon/c}{1/c^2-1/\kappa^2}\;,
\label{p0}
\end{equation}
and a modification of the celebrated equation $E_0=m\,c^2$ into
\begin{equation}
E_0=\frac{m\,c^2}{1-c^2/\kappa^2}\;.
\label{Emc2}
\end{equation}
Note that by Eqs.~(\ref{p0}) and (\ref{Emc2}) the right-hand side of the differential equation (\ref{eqforLfinal}) vanishes, so these choices for $p_0$ and $E_0$ are equivalent to setting the constants $\alpha'(0)$ and $\beta'(0)$ in Eq.~(\ref{eqforL'}) equal to zero.  Of course, Eq.~(\ref{Emc2}) is physically sound only if one can argue that, for a system of particles, the total energy in the centre-of-momentum frame stands in the same relation with the mass of the system, regarded as a single unit.  This is shown in Appendix~\ref{subsec:appendixB}.

Finally, let us consider the extreme cases of strong conventional anisotropy,  $\varepsilon=\pm 1$.  Now one of the two invariant speeds $c_+$ and $c_-$ is infinite, and it is convenient to write, as we already did in Sec.~\ref{subsubsec:1}, $C=c_+$ or $C=c_-$, according to whether $b$ is positive or negative, respectively. This amounts to setting $b=\varepsilon/C$, or equivalently $c=2\,C$.  The expressions for the basic dynamical quantities can either be obtained directly from Eqs.~(\ref{Lknotc})--(\ref{Eknotc}), or from Eqs.~(\ref{LEinst})--(\ref{EEinst}) with the replacements $\varepsilon=\pm 1$ and $c=2\,C$:
\begin{eqnarray}
L(u)=&-&\frac{m}{1/4C^2-1/\kappa^2}\left(1
+\frac{\varepsilon\,u}{C}\right)^{1/2-\varepsilon\,C/\kappa}\nonumber\\
&+&\left(p_0+\frac{m}{\varepsilon/2C+1/\kappa}\right)u
-\left(E_0-\frac{m}{1/4C^2-1/\kappa^2}\right)\;;
\label{Llim}
\end{eqnarray}
\begin{equation}
p(u)=-\frac{m}{\varepsilon/2C+1/\kappa}\left(1
+\frac{\varepsilon\,u}{C}\right)^{-1/2-\varepsilon\,C/\kappa}
+p_0+\frac{m}{\varepsilon/2C+1/\kappa}\;;
\label{plim}
\end{equation}
\begin{equation}
E(u)=\frac{m}{1/4C^2-1/\kappa^2}\left(1
+\frac{\varepsilon\,u}{C}\right)^{-1/2-\varepsilon\,C/\kappa}
\left(1+\left(\frac{\varepsilon}{2\,C}+\frac{1}{\kappa}\right)u\right)
+E_0-\frac{m}{1/4C^2-1/\kappa^2}\;.
\label{Elim}
\end{equation}
%

\subsubsection{Anisotropic Newtonian dynamics}
\label{subsec:3}

\begin{figure}[htbp]
\vbox{ \hfil
\scalebox{0.32}{{\includegraphics{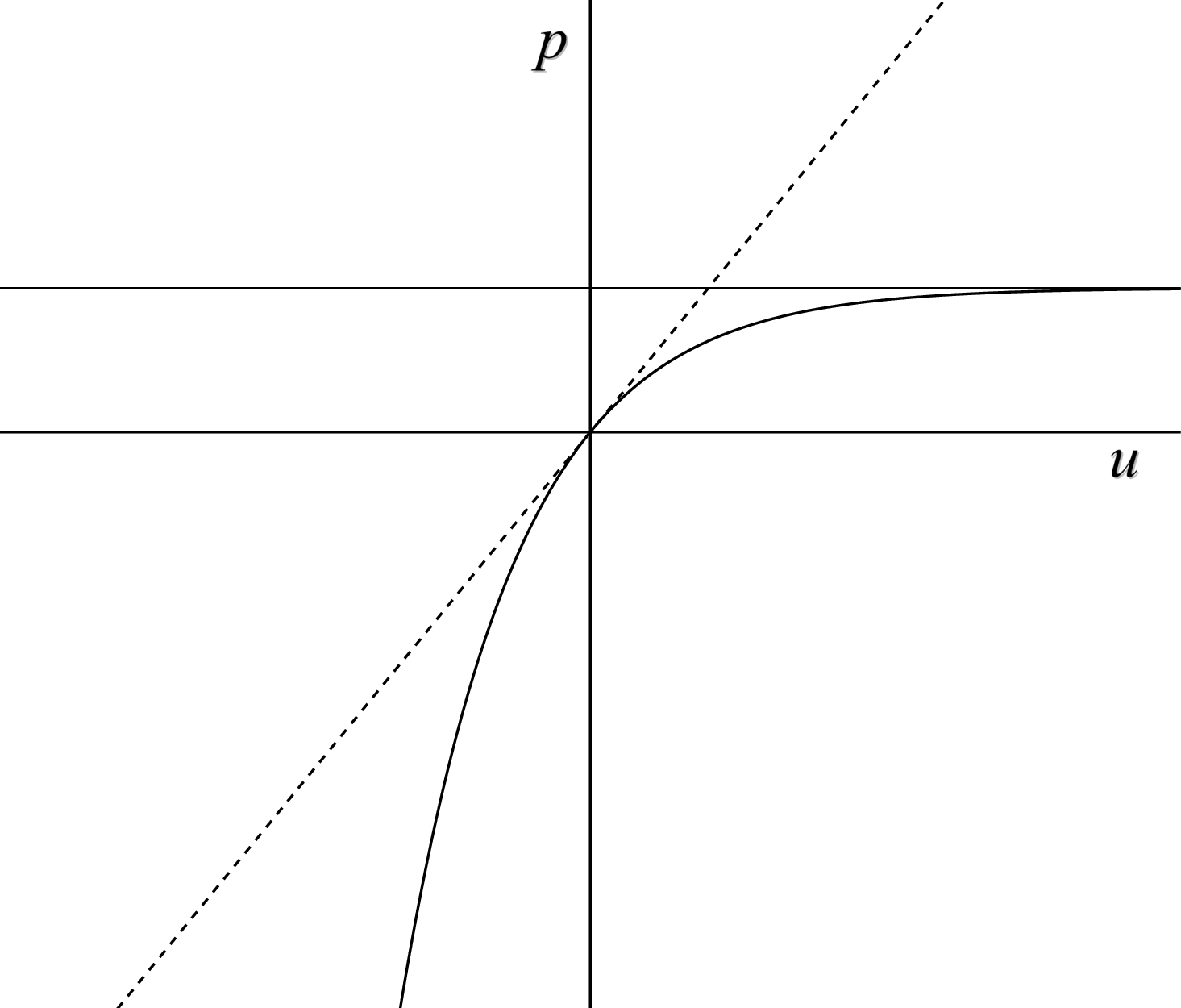}},\hskip1.5cm{\includegraphics{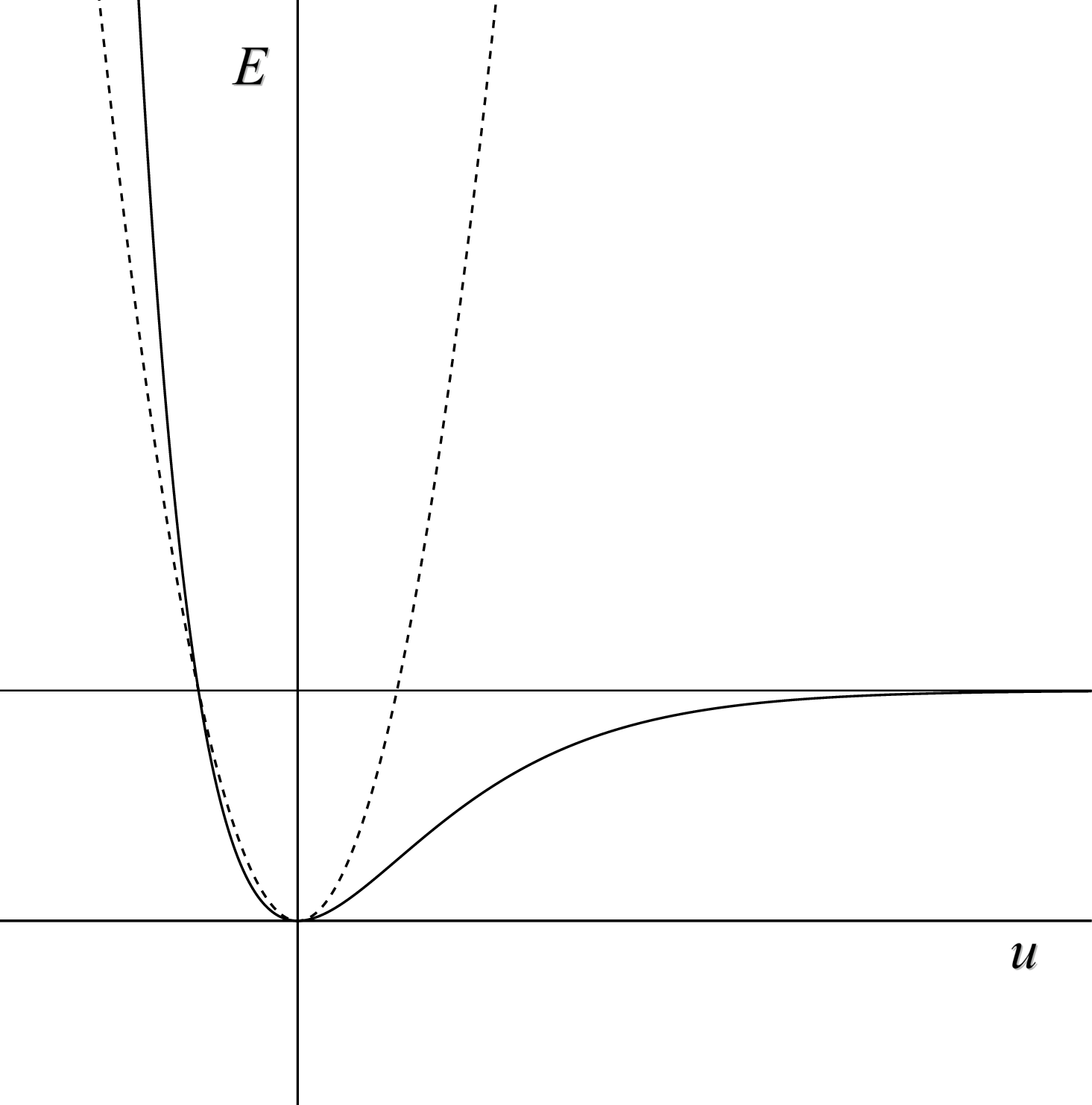}}}
\hfil }
\vskip.25cm
\caption{\small Comparison between the expressions given by Eqs.~(\ref{pnewt}) and (\ref{Enewt}) for $0<\kappa<+\infty$ (solid lines) and $|\kappa|=+\infty$ (dashed lines).  In both cases, the values of the quantities at rest have been chosen equal to zero. The plot on the left represents momentum, the one on the right represents energy.  Note the horizontal asymptotes of $p$ and $E$ as $u\to +\infty$ (displayed as thin solid straight lines).}
\label{fig2}
\end{figure}

We now consider the case $\sigma=\varepsilon=0$ (or, alternatively, $c=+\infty$).  On replacing $a=b=0$, $\mu=1/\kappa$, $h(u)=u$ into Eqs.~(\ref{Lknotc})--(\ref{Eknotc}), one finds:
\begin{equation}
L(u)=-m\,\kappa^2\left(1-\ee^{-u/\kappa}\right)
+\left(p_0+m\,\kappa\right)u-E_0\;;
\label{Lnewt}
\end{equation}
\begin{equation}
p(u)=m\,\kappa\,\left(1-\ee^{-u/\kappa}\right)+p_0\;;
\label{pnewt}
\end{equation}
\begin{equation}
E(u)=m\,\kappa^2\,\left(1-\ee^{-u/\kappa}\right)
-m\,\kappa\,u\,\ee^{-u/\kappa}+E_0\;.
\label{Enewt}
\end{equation}
Note that for $|u|\ll |\kappa|$ one recovers the usual expressions of isotropic Newtonian dynamics (apart from the constants $p_0$ and $E_0$).  Figure~\ref{fig2} shows a comparison between these expressions and those valid assuming isotropy.  Both $p$ and $E$ tend to asymptotic values as $u$ tends to $+\infty$ if $\kappa>0$, or $-\infty$ if $\kappa<0$.  This behaviour, although mathematically interesting, is however irrelevant from a physical point of view, because it concerns the high-speed regime of a theory which is, in fact, only a low-speed approximation.

Equations~(\ref{Lnewt})--(\ref{Enewt}) can be obtained, formally, as the limit for $c\to +\infty$ of the corresponding expressions in anisotropic Einstein's dynamics, Eqs.~(\ref{LEinst})--(\ref{EEinst}).  However, if the limit is taken {\em after\/} the choices (\ref{p0}) and (\ref{Emc2}) for $E_0$ and $p_0$ are made, one gets a rather bizarre behaviour for the rest energy and momentum in the Newtonian limit, as these quantities would diverge as $|\kappa|\to +\infty$.  Of course, in drawing this unphysical consequence one ignores the fact that in the real world $c$ is finite, and that experiments give $|\kappa|>c$, so one cannot really study the case $c\to +\infty$ keeping $|\kappa|$ finite.

\subsection{Case $|\kappa|=c$}
\label{subsec:k=c}

The exceptional cases with $\kappa=\pm c$ have of course no physical interest; nevertheless, we present the results for the sake of completeness.  The solution of Eq.~(\ref{eqforLfinal}) can be written in a unified way as
\begin{equation}
L(u)=\frac{m\,c^2}{4}\left(1-\left(\frac{1}{\kappa}-\frac{\varepsilon}{c}\right)u\right)
\ln\left(\frac{1-\left(1/\kappa-\varepsilon/c\right)u}{1+\left(1/\kappa
+\varepsilon/c\right)u}\right)+\left(p_0+m\,\kappa/2\right)u-E_0\;,
\label{Lex}
\end{equation}
where again Eqs.~(\ref{ab}) and (\ref{muu}) have been used, and $m=L''(0)$.  The momentum and energy are then, respectively:
\begin{equation}
p(u)=\frac{m\,\kappa}{2}\frac{\left(1/\kappa+\varepsilon/c\right)u}{1
+\left(1/\kappa+\varepsilon/c\right)u}-\frac{m\,c^2}{4}\left(\frac{1}{\kappa}-\frac{\varepsilon}{c}\right)
\ln\left(\frac{1-\left(1/\kappa-\varepsilon/c\right)u}{1+\left(1/\kappa
+\varepsilon/c\right)u}\right)+p_0\;;
\label{pex}
\end{equation}
\begin{equation}
E(u)=-\frac{m\,\kappa\,u}{2}\frac{1}{1+\left(1/\kappa+\varepsilon/c\right)u}-\frac{m\,c^2}{4}\ln\left(\frac{1-\left(1/\kappa-\varepsilon/c\right)u}{1+\left(1/\kappa+\varepsilon/c\right)u}\right)+E_0\;.
\label{Eex}
\end{equation}
Note that these expressions cannot be obtained by taking the limits $\kappa\to\pm c$ of the corresponding expressions in Sec.~\ref{subsec:kneqc}, because such limits do not exist.

In the extreme cases of strong conventional anisotropy ($\varepsilon=\pm 1$, $c=2\,C$), these expressions reduce to:
\begin{equation}
L(u)=-\frac{\varepsilon\,m\,\kappa\,C}{2}\left(1-\left(\frac{1}{\kappa}-\frac{\varepsilon}{2C}\right)u\right)\ln\left(1+\frac{\varepsilon\,u}{C}\right)+\left(p_0+m\,\kappa/2\right)u-E_0\;;
\label{Lex!}
\end{equation}
\begin{equation}
p(u)=\frac{m\,\kappa}{2}\frac{\left(1/\kappa+\varepsilon/2C\right)u}{1
+\left(1/\kappa+\varepsilon/2C\right)u}+\frac{\varepsilon\,m\,\kappa\,C}{2}\left(\frac{1}{\kappa}-\frac{\varepsilon}{2C}\right)
\ln\left(1+\frac{\varepsilon\,u}{C}\right)+p_0\;;
\label{pex!}
\end{equation}
\begin{equation}
E(u)=-\frac{m\,\kappa\,u}{2}\frac{1}{1+\left(1/\kappa+\varepsilon/2C\right)u}+\frac{\varepsilon\,m\,\kappa\,C}{2}\ln\left(1+\frac{\varepsilon\,u}{C}\right)+E_0\;.
\label{Eex!}
\end{equation}
%

\section{Hamiltonian and dispersion relation}
\label{sec:hamiltonian}
\setcounter{equation}{0}

Comparing Eq.~(\ref{dT}) with Hamilton's equation
\begin{equation}
u=\frac{\d H(p)}{\d p}\;,
\label{true-ham}
\end{equation}
allows us to identify the Hamiltonian $H(p)$ as $E(u(p))$, up to a velocity-independent additive term that we set equal to zero.  (Of course, the same expression for $H$ can be obtained as the Legendre transform of $L$.)  

In principle, one could find $H$ directly, solving a differential equation  as we did for $L$ in Sec.~\ref{sec:examples}.  Indeed, by a further differentiation of Hamilton's equation (\ref{true-ham}) we get
\begin{equation}
\d u=\frac{\d^2 H}{\d p^2}\,\d p\;.
\label{du}
\end{equation}
Combining this with Eq.~(\ref{momentum}) for the relevant case $\lambda=1$, we arrive at the following differential equation for $H(p)$: 
\begin{equation}
\left(p-\mu\,H-\nu\right)\frac{\d^2 H}{\d p^2}=\frac{\d H}{\d
p}\,\varphi\left(\frac{\d H}{\d p}\right)\;.
\label{diff-disp}
\end{equation}
In general, however, Eq.~(\ref{diff-disp}) is not easy to solve.  Therefore, it is more convenient to rely on the expressions for $E(u)$ and $p(u)$, trying to eliminate $u$ to obtain an implicit relation between energy and momentum, from which the Hamiltonian can, in principle, be extracted by local inversion.  This implicit relation is the particle version of a dispersion relation. 

\subsection{Case $|\kappa|\neq c$}
\label{subsubsec:kappaneqcham}

\subsubsection{Anisotropic Einstein's dynamics}
\label{subsec:1ham}

\begin{figure}[htbp]
\vbox{ \hfil
\scalebox{0.35}{{\includegraphics{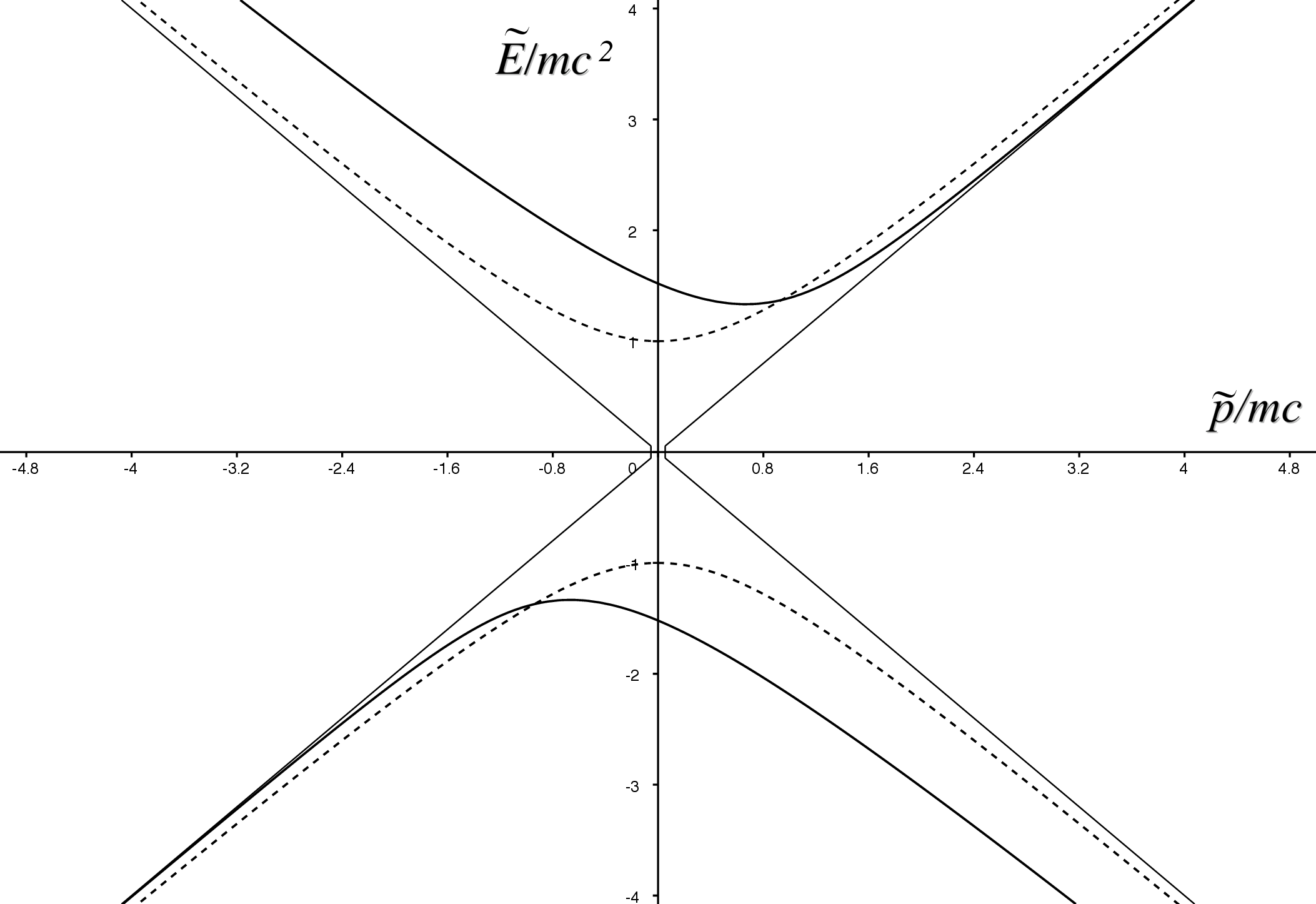}}}
\hfil }
\vskip.25cm
\caption{\small Comparison between the set of pairs $(\tilde{p},\widetilde{E})$ that satisfy Eq.~(\ref{disp-Einst!}) for $\kappa=2\,c$ (thick solid line) and $|\kappa|=+\infty$ (dashed line).  The conventional anisotropy parameter $\varepsilon$ has been set equal to zero, and the asymptotes are also displayed (thin solid straight lines).}
\label{fig3}
\end{figure}

It is convenient to define the new quantities
\begin{equation}
\tilde{p}(u):=p(u)-p_0+m\,\frac{1/\kappa-\varepsilon/c}{1/c^2-1/\kappa^2}
\label{ptilde}
\end{equation}
and
\begin{equation}
\widetilde{E}(u):=E(u)-E_0+\frac{m}{1/c^2-1/\kappa^2}\;,
\label{Etilde}
\end{equation}
which are such that
\begin{equation}
\tilde{p}(0)=m\,\frac{1/\kappa-\varepsilon/c}{1/c^2-1/\kappa^2}
\label{ptilde0}
\end{equation}
and
\begin{equation}
\widetilde{E}(0)=\frac{m}{1/c^2-1/\kappa^2}\;,
\label{Etilde0}
\end{equation}
independently of the choices for $p_0$ and $E_0$.  (Hence, $\tilde{p}$ and $\widetilde{E}$ coincide with momentum and energy once the natural choices for $p_0$ and $E_0$ have been made, as in Eqs.~(\ref{p0}) and (\ref{Emc2}).)  Then, let us form their linear combinations
\begin{equation}
\left(1-\varepsilon\right)\widetilde{E}-\tilde{p}\,c\;,\qquad\qquad
\left(1+\varepsilon\right)\widetilde{E}+\tilde{p}\,c\;,
\label{+-}
\end{equation}
and multiply and divide these by each other.  In this way, we obtain two expressions that both contain $u$ in the combination
\[ \frac{1-\left(1-\varepsilon\right)u/c}{1+\left(1
+\varepsilon\right)u/c}\;,\]
so at the end we can form a $u$-independent expression:
\begin{equation}
\left(\left(1-\varepsilon\right)\widetilde{E}-\tilde{p}\,c\right)^{1-c/\kappa}
\left(\left(1+\varepsilon\right)\widetilde{E}+\tilde{p}\,c\right)^{1+c/\kappa} =\frac{m^2\,c^2}{\left(1/c-1/\kappa\right)^{1+c/\kappa}
\left(1/c+1/\kappa\right)^{1-c/\kappa}}\;.
\label{disp-Einst}
\end{equation}
An alternative form of this equation, which allows also for negative values of $\widetilde {E}$, is
\begin{equation}
\left(\left(1-\varepsilon\right)\widetilde{E}-\tilde{p}\,c\right)
\left(\left(1+\varepsilon\right)\widetilde{E}+\tilde{p}\,c\right)^{\frac{1+c/\kappa}{1-c/\kappa}} =\frac{m^{\frac{2}{1-c/\kappa}}\,c^{\frac{2}{1-c/\kappa}}}{\left(1/c-1/\kappa\right)^{\frac{1+c/\kappa}{1-c/\kappa}}
\left(1/c+1/\kappa\right)}\;.
\label{disp-Einst!}
\end{equation}
This is the sought-for dispersion relation, that generalises the well-known equation $E^2-p^2\,c^2=m^2\,c^4$ to the case in which space is anisotropic.  Figure~\ref{fig3} shows a comparison between these two expressions.

To the first order in $c/\kappa$, Eq.~(\ref{disp-Einst}) leads to the approximate relation
\begin{equation}
\left(\left(1-\varepsilon\right)\widetilde{E}-\tilde{p}\,c\right)
\left(\left(1+\varepsilon\right)\widetilde{E}+\tilde{p}\,c\right)
\approx m^2\,c^4\left(1+\frac{c}{\kappa}\ln\left(\frac{\left(1
-\varepsilon\right)\widetilde{E}-\tilde{p}\,c}{\left(1
+\varepsilon\right)\widetilde{E}+\tilde{p}\,c}\right)\right)\;, 
\label{disp-Einst-app}
\end{equation}
which can be used in order to place experimental bounds on $c/\kappa$.

In the cases of strong conventional anisotropy ($\varepsilon=\pm 1$), using the same notations as in the final part of Sec.~\ref{subsec:1} and eliminating $u$, we can directly write the Hamiltonian:
\begin{equation}
\widetilde{H}(\tilde{p})=-\varepsilon\,\tilde{p}\,C\left(1+\frac{\varepsilon/2C+1/\kappa}{\varepsilon/2C-1/\kappa}\left(\frac{m}{\left(\varepsilon/2C+1/\kappa\right)\tilde{p}}\right)^\frac{2}{1+2\varepsilon C/\kappa}\right)\;.
\label{dispC}
\end{equation}
Of course, this can also be recovered by Eq.~(\ref{disp-Einst!}) with the appropriate substitutions.

\subsubsection{Anisotropic Newtonian dynamics}
\label{subsec:3ham}

\begin{figure}[htbp]
\vbox{ \hfil
\scalebox{0.35}{{\includegraphics{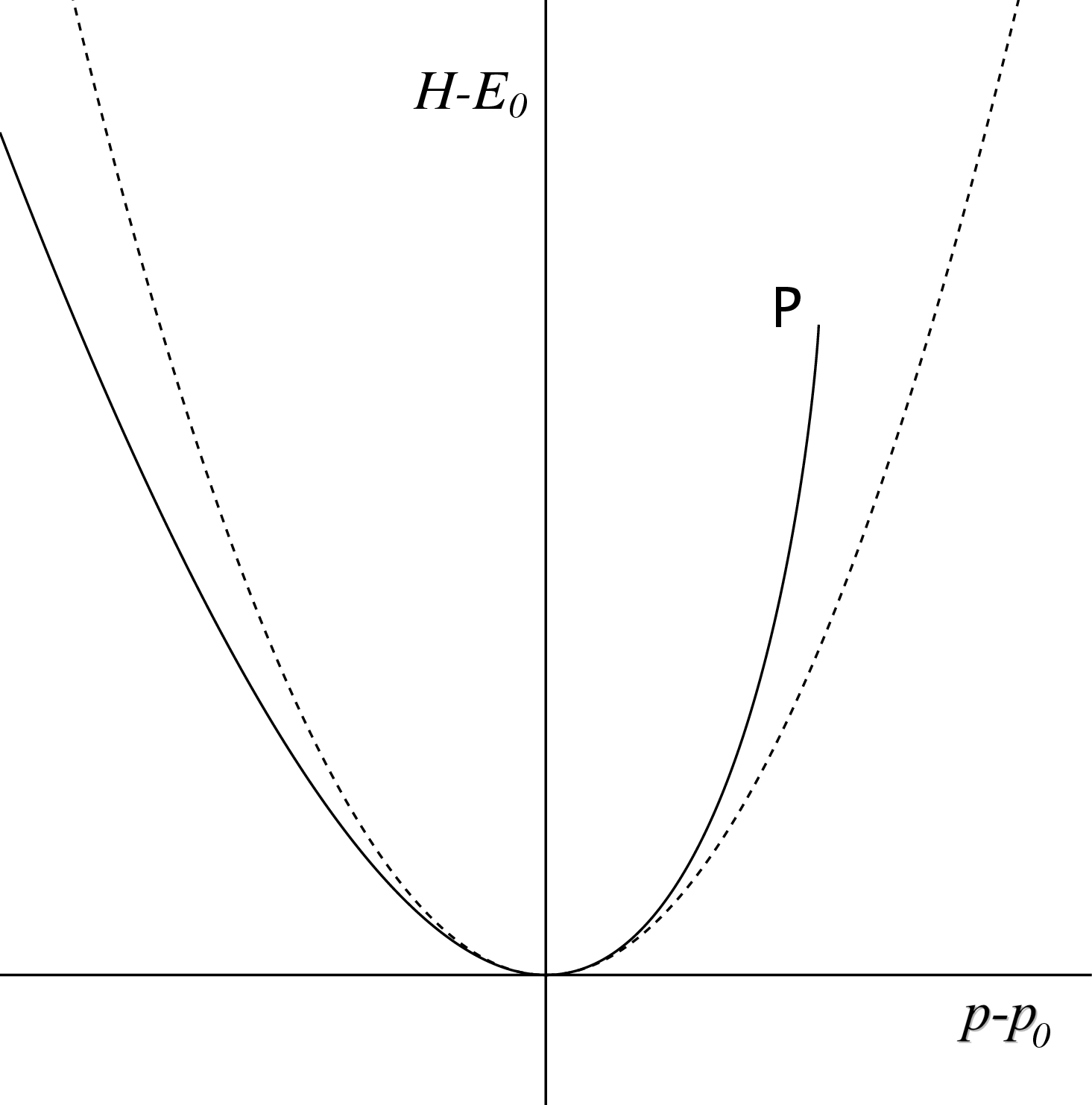}}}
\hfil }
\vskip.25cm
\caption{\small Comparison between the set of pairs $(p-p_0,H-E_0)$ that satisfy Eq.~(\ref{HN}) for $0<\kappa<+\infty$ (solid line) and $|\kappa|=+\infty$ (dashed line).  Note that the curve for a finite $\kappa$ possesses an end-point ({\sf P} in the diagram), which corresponds to the finite asymptotic values of $p$ and $E$ as $u\to +\infty$.}
\label{fig4}
\end{figure}

The Hamiltonian is
\begin{eqnarray}
H(p)&=&E_0+\kappa\,\left(p-p_0\right)+m\,\kappa^2\left(1
-\frac{p-p_0}{m\,\kappa}\right)\ln\left(1-\frac{p-p_0}{m\,\kappa}\right)\nonumber\\
&=&E_0+\frac{1}{2\,m}\left(p-p_0\right)^2+{\cal O}\left((p-p_0)^3\right)\;.
\label{HN}
\end{eqnarray}
Figure~\ref{fig4} shows a comparison between the cases with and without anisotropy.
%

\subsection{Case $|\kappa|=c$}
\label{subsubsec:kappa=0ham}

By the same procedure as in the previous section we find
\begin{equation}
\left(\frac{1}{\kappa}-\frac{\varepsilon}{c}\right)\widetilde{E}-\tilde{p}
=-\frac{m\,u}{1+\left(1/\kappa+\varepsilon/c\right)u}
\label{-}
\end{equation}
and
\begin{equation}
\left(\frac{1}{\kappa}+\frac{\varepsilon}{c}\right)\widetilde{E}+\tilde{p}
=-\frac{m\,c^2}{2\,\kappa}\ln\left(\frac{1-\left(1/\kappa-\varepsilon/c\right)u}{1+\left(1/\kappa
+\varepsilon/c\right)u}\right)\;,
\label{+}
\end{equation}
where now $\widetilde{E}=E-E_0$ and $\tilde{p}=p-p_0$.  Extracting $u$ from Eq.~(\ref{+}) and replacing it into Eq.~(\ref{-}) one finds, after trivial algebra, the dispersion relation:
\begin{equation}
\left(\frac{1}{\kappa}-\frac{\varepsilon}{c}\right)\widetilde{E}-\tilde{p}
+\frac{m\,\kappa}{2}-\frac{m\,\kappa}{2}\exp\left(-\frac{2\,\kappa}{m\,c^2}\left(\left(\frac{1}{\kappa}
+\frac{\varepsilon}{c}\right)\widetilde{E}+\tilde{p}\right)\right)=0\;.
\label{disp-except}
\end{equation}

For strong conventional anisotropy we find
\begin{equation}
\left(\frac{1}{\kappa}-\frac{\varepsilon}{2C}\right)\widetilde{E}-\tilde{p}
+\frac{m\,\kappa}{2}-\frac{m\,\kappa}{2}\exp\left(-\frac{\kappa}{2\,m\,C^2}\left(\left(\frac{1}{\kappa}
+\frac{\varepsilon}{2C}\right)\widetilde{E}+\tilde{p}\right)\right)=0\;,
\label{disp-except!}
\end{equation}
which can also be obtained by the straightforward substitution $c=2C$ into Eq.~(\ref{disp-except}).

\section{Geometrical formulation}
\label{sec:geometry}
\setcounter{equation}{0}

Special relativistic kinematics can be given a geometric interpretation introducing the Minkowski quadratic form
\begin{equation}
\eta(\d t,\d x)=\d t^2-\d x^2/c^2\;,
\label{mink}
\end{equation}
which is invariant under Lorentz transformations of the coordinates.  We now ask what is the most general function of $\d t$ and $\d x$ that is invariant under the transformations found in Sec.~\ref{sec:kin}.  That is, we look for a function $\zeta$ such that
\begin{equation}
\zeta(\d\bar{t},\d\bar{x})=\zeta(\d t,\d x)\,,\qquad\forall v\in J\;.
\label{F}
\end{equation}
Differentiating Eq.~(\ref{F}) with respect to $v$ and evaluating the result for $v=0$ we find
\begin{equation}
\left(\left(\frac{1}{\kappa}-\frac{b}{2}\right)\d t+a\,\d
x\right)\frac{\partial\,\zeta}{\partial\,\d t} +\left(\d
t+\left(\frac{1}{\kappa}
+\frac{b}{2}\right)\d x\right)\frac{\partial\,\zeta}{\partial\,\d x}=0\;.
\label{diffeqf}
\end{equation}
This is a particular case of the partial differential equation solved in Appendix~\ref{subsec:appendixC}, corresponding to: $X=\d t$; $Y=\d x$; $A=1/\kappa-b/2$; $B=a$; $C=1$; $D=1/\kappa+b/2$; $F=\zeta$; $\Psi=\varphi$; $\Theta=h$.  Hence, 
\begin{equation}
\zeta(\d t,\d x)=\ee^{-2h(\d x/\d t)/\kappa}\left(\d t^2+b\,\d t\,\d x-a\,\d x^2\right)\;,
\label{invariant}
\end{equation}
where the expression (\ref{phiab'}) for $\varphi$ has been used, is the extension of the quadratic form (\ref{mink}).  

Anisotropic kinematics can then be interpreted in terms of a spacetime geometry by introducing the pseudo-norm
\begin{equation}
F(\xi)=\ee^{-h(\xi^1/\xi^0)/\kappa}\left((\xi^0)^2+b\,\xi^0\,\xi^1-a\,(\xi^1)^2\right)^{1/2}\;,
\label{norm-gen}
\end{equation}
where $\xi$ is a generic vector in spacetime.  The most general spacetime structure compatible with the relativistic framework considered in this paper is thus a pseudo-Finslerian, not a pseudo-Riemannian,\footnote{The difference between Finslerian~\cite{finsler} and pseudo-Finslerian~\cite{pseudo-finsler} structures on a manifold is the same as between Riemannian and pseudo-Riemannian ones.} one.  The pseudo-Finslerian character, associated with the exponential factor in Eqs.~(\ref{invariant}) and (\ref{norm-gen}), is unavoidable when one wants to allow for a mechanical anisotropy, through the parameter $\kappa$.  It is interesting to note that the other kind of anisotropy, linked to the parameter $\varepsilon$, does not alter the pseudo-Riemannian character of spacetime.  This can be exemplified in the physically interesting case of anisotropic Einstein's kinematics, where Eq.~(\ref{invariant}) becomes\footnote{In the case $c_+=c_-$ ({\em i.e.\/}, $\varepsilon=0$), this expression was also considered in  Refs.~\cite{bogoslovsky, bogo, budden}.}
\begin{equation}
\zeta(\d t,\d x)=\left(\d t+\d x/c_+\right)^{1-c/\kappa}\left(\d t-\d
x/c_-\right)^{1+c/\kappa}\;.
\label{finsler}
\end{equation}
When $|\kappa|=+\infty$ but $\varepsilon\neq 0$, this is just the Minkowski quadratic form (\ref{mink}) in non-Lorentzian rectilinear coordinates, defined operationally through a non-standard synchronisation procedure.  The fact that the value of $\varepsilon$ does not alter the spacetime structure agrees with the thesis that the anisotropy associated with such a parameter is an artifact of a convention rather than a physical feature (see the discussion in Sec.~\ref{subsubsec:anisotropy}).  Indeed, one can always eliminate such an anisotropy by synchronising clocks according to the Einstein procedure, which amounts to choosing Lorentzian coordinates in spacetime.  On the other hand, a finite value of $\kappa$ implies true physical effects (anisotropic time dilation and length contraction, modified dispersion relation, ...), that cannot be gauged away by a stipulation, just as the pseudo-Finslerian character of spacetime cannot be reduced to a pseudo-Riemannian one by a suitable choice of coordinates.

For a particle tracing out a differentiable worldline $x^a(\theta)$ in spacetime,\footnote{The indices $a$, $b$, $\ldots$ run from 0 to 1.} where $\theta$ is an arbitrary parameter, we can define the tangent vector with components $U^a=\d x^a/\d\theta$, and the proper time $\tau$ such that $\d\tau=F(U)\,\d\theta$.  Of course, $U^1=u\,U^0$, so by a comparison between Eqs.~(\ref{Lknotc}) and (\ref{norm-gen}) we note that for $|\kappa|\neq c$, 
\begin{equation}
L(u)\,\d t=-\frac{m}{1/c^2-1/\kappa^2}\,F(U)\,\d\theta+\left(\left(p_0-m\,\frac{1/\kappa-\varepsilon/c}{1/c^2
-1/\kappa^2}\right)U^1
-\left(E_0-\frac{m}{1/c^2-1/\kappa^2}\right)U^0\right)\d\theta\;.
\label{laggeom}
\end{equation}
Hence, apart from a constant additive term,\footnote{Which vanishes when the rest momentum and energy are given by the expressions (\ref{p0}) and (\ref{Emc2}).} the action turns out to be proportional to the particle proper time, as it happens in ordinary (isotropic) special relativity.  

Two comments are in order, of opposite flavour.  First, we see that the Newtonian Lagrangian $mu^2/2$, in spite of its appearance, is related to spacetime geometry.  For it is the limit, when $|\kappa|\to +\infty$, of the Lagrangian (\ref{Lnewt}), which is the particular case of Eq.~(\ref{laggeom}) corresponding to the norm
\begin{equation}
F(\xi)=\exp\left(-\frac{1}{\kappa}\,\frac{\xi^1}{\xi^0}\right)|\xi^0|
\label{dtaunewt}
\end{equation}
and to $c=+\infty$.  Second, although it is customary in the literature about relativity to choose a Lagrangian such that the corresponding action turns out to be related to the proper time, this should not be taken as a dogma.  In the exceptional cases with $\kappa=\pm c$, the pseudo-norm is
\begin{equation}
F(\xi)=\left|\xi^0-\left(\frac{1}{\kappa}-\frac{\varepsilon}{c}\right)\xi^1\right|\;,
\label{tauexe}
\end{equation}
so $F(U)\,\d\theta=\left(1-\left(1/\kappa-\varepsilon/c\right)u\right)\d t$, which does not contribute to the equation of motion.  Hence, not only Eq.~(\ref{laggeom}) does not apply to these cases, but {\em no\/} term proportional to proper time in the action leads to a non-trivial dynamics --- in fact, Eq.~(\ref{Lex}) does not appear to have a geometrical interpretation. 

Equation (\ref{laggeom}) allows one to define a Lagrangian in spacetime,
\begin{equation}
{\cal L}(U)=-\frac{m}{1/c^2-1/\kappa^2}\,F(U)+\left(\left(p_0-m\,\frac{1/\kappa-\varepsilon/c}{1/c^2
-1/\kappa^2}\right)U^1
-\left(E_0-\frac{m}{1/c^2-1/\kappa^2}\right)U^0\right)\;,
\label{lagst}
\end{equation}
so one can define a momentum one-form in spacetime, with components $p_a=\partial {\cal L}/\partial U^a$.  It is easy to see that $p_0=-E(u)$ and $p_1=p(u)$.  Note that, since ${\cal L}$ is a homogeneous function of degree 1, the components $p_a$ are independent of the parametrisation of the particle worldline.  It is also worth stressing that the relation between $p_a$ and $U^a$ is not trivial.  Indeed, on defining the quantities $\tilde{p}_0=-\widetilde{E}$ and $\tilde{p}_1=\tilde{p}$ as we did in Sec.~\ref{sec:hamiltonian}, one finds
\begin{equation}
\tilde{p}_a=-\frac{m}{1/c^2-1/\kappa^2}\,\frac{\partial F(U)}{\partial U^a}=-\frac{m}{1/c^2-1/\kappa^2}\,\frac{\g_{ab}(U)\,U^b}{F(U)}\;,
\label{Pa}
\end{equation}
where in the last equality we have used Euler's theorem for the function  $\partial F(U)^2/\partial U^a$, which is homogeneous of degree 1, and  the definition~\cite{finsler,pseudo-finsler,gls}
\begin{equation}
\g_{ab}(U)=\frac{1}{2}\,\frac{\partial^2 F(U)^2}{\partial U^a\,\partial U^b}\;.
\label{metric}
\end{equation}
It is now possible, introducing coefficients $\g^{ab}(\tilde{p})$ such that\footnote{The $\g^{ab}$ are homogeneous functions of degree 0 of their arguments (as the $\g_{ab}$ are), so their definition is insensitive to the coefficients in Eq.~(\ref{Pa}).} $\g^{ab}(\tilde{p})\,\g_{bc}(U)={\delta^a}_c$, to write the dispersion relations of Sec.~\ref{subsubsec:kappaneqcham} in the covariant form~\cite{gls}
\begin{equation}
\g^{ab}(\tilde{p})\,\tilde{p}_a\,\tilde{p}_b=\frac{m^2}{\left(1/c^2-1/\kappa^2\right)^2}\;,
\label{dispcov}
\end{equation}
where the relation $\g_{ab}(U)\,U^a\,U^b=F(U)^2$, which follows from the fact that $F^2$ is homogeneous of degree 2, has been used.

\section{Comments}
\label{sec:comments}
\setcounter{equation}{0}

In this paper we have developed the foundations of particle dynamics for theories that are compatible with the relativity principle and include a mechanical anisotropy.  We have seen that anisotropy can be taken into account through two parameters, one of which ($\varepsilon$) has a merely conventional status, while the other ($\kappa$) accounts for real physical effects.  It may then seem odd, that the parameter $\varepsilon$ enters in the expressions of physically relevant quantities such as momentum and energy; for this appears to offer an opportunity to test conventional isotropy by means of dynamical experiments, which should of course be impossible.  In fact, the value of $\varepsilon$ cannot be dynamically determined, independently of kinematical considerations, because any such experiment requires that one could also determine velocities, which again presupposes a synchronisation procedure.  Thus, the presence of $\varepsilon$ into equations such as (\ref{pEinst}) and (\ref{EEinst}) is no longer surprising, its role being analogous to that played by a conversion factor.

On the contrary, the value of $\kappa$ {\em can\/} be determined experimentally.  For example, one might study momentum conservation in a collision.  Having synchronised clocks with some procedure (corresponding to a value for $\varepsilon$) and defined velocities correspondingly, one can look for the value of $\kappa$ that provides a best fit when used in Eq.~(\ref{pEinst}).  Another possibility, less direct but perhaps more practicable, is to place constraints on the deviations from standard dispersion relations, as it is now fashionable to do within the context of the so-called ``quantum gravity phenomenology''~\cite{mattingly}.

We have deliberately restricted our investigation to those cases in which time and space are homogeneous, and space is geometrically homogeneous and isotropic; the only manifestations of anisotropy being mechanical.  This choice is motivated by the desire to set up a kinematical framework based on reference frames of the type usually considered in Newtonian mechanics and special relativity.  Of course, it would be interesting to generalise mechanics further, perhaps even contemplating situations in which space and time are inhomogeneous, possibly at short scales.  (This is what several people believe might happen in a more sophisticated theory of spacetime structure.)  However, such generalisations must necessarily proceed along lines that differ from those of the present paper.  For example, one might explore anisotropic deformations of de Sitter relativity~\cite{deSitter}, or try to include anisotropy at the level of the algebra of generators~\cite{bacry}.

Recently, motivated by experimental data about ultra-high-energy cosmic rays~\cite{sigl}, there have been some suggestions to modify the standard dispersion relations in such a way that the relativity principle is preserved, allowing at the same time for the existence of an invariant energy scale.  Such modifications have been developed  essentially within two theoretical frameworks: Nonlinear representations of the Lorentz group~\cite{dsr}, and the so-called $\kappa$-deformations of it~\cite{noncommut} (see also~\cite{DSRdeSitter}).  In both cases, the emphasis is on energy-momentum space, while the role of configuration variables, such as position and velocity, remains somewhat unclear~\cite{position}.  Of course, the corresponding  dispersion relations differ from those found in Sec.~\ref{sec:hamiltonian}; hence, these frameworks are incompatible with at least one of the hypotheses on time and space made in the present derivation.  (It is also possible that they are not fundamental as suggested recently~\cite{meas}.)  Another, less fancy, origin for modified dispersion relations could simply be a breaking of Lorentz invariance, perhaps combined with small-scale inhomogeneity.  This happens, {\em e.g.\/}, if one replaces space by a discrete structure, in which case Lorentz invariance is recovered only at low momenta, which do not probe the lattice structure of the background and thus the fundamental asymmetry between space and time.  A similar behaviour is exhibited by condensed matter models, such as Bose-Einstein condensates~\cite{matt}.

Using the equivalence principle in order to argue that gravitational effects can be locally gauged away, one can construct a theory of curved pseudo-Finslerian spacetime that locally reduces to the one of Eq.~(\ref{finsler}), thus obtaining an extension of general relativity.  One might even construct a theory in which the amount of anisotropy changes from place to place, {\em e.g.\/}, replacing the constant parameter $\kappa$ by a field~\cite{bogo}.  In this case, it is possible to  envisage situations in which spacetime here and now is almost Lorentzian, whereas elsewhere (perhaps in regions of strong gravity, or in the early universe) it is highly non-pseudo-Riemannian.

\section*{Acknowledgements}

It is a pleasure to thank Jean-Marc L\'evy-Leblond and Abraham Ungar for correspondence, and an anonymous referee for suggesting a correction.  S.S.\ is grateful to Stefano Liberati, Lorenzo Sindoni and Matt Visser for stimulating discussions.

\renewcommand{\thesubsection}{\Alph{subsection}}
\section*{Appendices}
\subsection{Alternative derivation of the transformation law}
\label{subsec:appendixA}
\renewcommand{\theequation}{\thesubsection.\arabic{equation}}
\setcounter{equation}{0}

Replacing Eq.~(\ref{lambda}) and the corresponding expressions for $\Lambda(u)$ and $\Lambda(\Phi(u,v))$ into Eq.~(\ref{associative}) we find, after elementary manipulations:
\begin{equation}
A(v)A(u)\left(1+u\,\xi(v)\right)=A(\Phi(u,v))\;;
\label{A1}
\end{equation}
\begin{equation}
\Phi(u,v)=\frac{v+u\,\eta(v)}{1+u\,\xi(v)}\;;
\label{A2}
\end{equation}
\begin{equation}
\xi(u)+\xi(v)\,\eta(u)=\left(1+u\,\xi(v)\right)\xi(\Phi(u,v))\;;
\label{A3}
\end{equation}
\begin{equation}
v\,\xi(u)+\eta(u)\,\eta(v)=\left(1+u\,\xi(v)\right)\eta(\Phi(u,v))\;.
\label{A4}
\end{equation}
From Eq.~(\ref{A2}) we immediately get the expression (\ref{phiab'}) for $\varphi(u)$.  Differentiating Eqs.~(\ref{A3}) and (\ref{A4}) with respect to $v$ and evaluating the result for $v=0$, one finds:
\begin{equation}
a\,\eta(u)=a\,u\,\xi(u)+\left(1+b\,u-a\,u^2\right)\xi'(u)\;;
\label{A5}
\end{equation}
\begin{equation}
\xi(u)+b\,\eta(u)=a\,u\,\eta(u)+\left(1+b\,u-a\,u^2\right)\eta'(u)\;.
\label{A6}
\end{equation}
Replacing $\xi(u)$ from Eq.~(\ref{A6}) into Eq.~(\ref{A5}), we obtain $\eta''(u)=0$, which is trivially integrated with the conditions $\eta(0)=1$, $\eta'(0)=b$, to obtain Eq.~(\ref{eta}).  Equation (\ref{A6}) gives then immediately the expression (\ref{xi}) for $\xi(u)$.

The function $A$ is still undetermined, but we can now find its general form by replacing the expression (\ref{xi}) into Eq.~(\ref{A1}), then differentiating the resulting equation with respect to $v$, and finally evaluating the result for $v=0$.  One finds, remembering that $A(0)=1$,
\begin{equation}
A'(0)A(u)+a\,u\,A(u)=A'(u)\varphi(u)\;.
\label{FF'}
\end{equation}
This equation can be immediately integrated to obtain Eq.~(\ref{A!}).

\subsection{On the mass-energy relation}
\label{subsec:appendixB}
\renewcommand{\theequation}{\thesubsection.\arabic{equation}}
\setcounter{equation}{0}

Consider, in a reference frame $\cal K$, two particles with masses $m_1$ and $m_2$, and velocities $u_1$ and $u_2$, such that
\begin{equation}
p(m_1,u_1)+p(m_2,u_2)=p(M,0)=p_0(M)
\label{p+p}
\end{equation}
for some $M$, function of $m_1$, $m_2$, $u_1$, $u_2$.  Thus, $\cal K$ can be regarded as the centre-of-momentum frame for the system,\footnote{When $\varepsilon=0$ and $|\kappa|=+\infty$, the centre-of-momentum frame can equivalently be defined as the one where total momentum vanishes.  This is not the appropriate characterisation when anisotropy is present in some form, as one can realise considering a situation in which the system is made of a single particle.  Also, note that defining a centre-of-mass frame is problematic in mechanics where there is no absolute time~\cite{com}.} and $M$ as its total mass.  In a reference frame $\overline{\cal K}$, in which $\cal K$ moves with velocity $v$, we have
\begin{equation}
p(m_1,\Phi(u_1,v))+p(m_2,\Phi(u_2,v))=p(M,v)\;.
\label{p+p-bar}
\end{equation}
The arbitrariness of $v$ allows us to find the value of $M$.  Indeed, differentiating Eq.~(\ref{p+p-bar}) with respect to $v$, and setting $v=0$, we get
\begin{equation}
p'(m_1,u_1)\varphi(u_1)+p'(m_2,u_2)\varphi(u_2)=p'(M,0)=L''(M,0)=M\;,
\label{mass}
\end{equation}
where a prime denotes the derivative with respect to velocity.  Remembering now Eq.~(\ref{dT}), Eq.~(\ref{momentum}) with $\lambda=1$, and Eqs.~(\ref{nu}) and (\ref{beta'}), we find
\begin{equation}
p'(m,u)\varphi(u)=\frac{1}{u}\left(p(m,u)-p_0(m)-\mu\left(E(m,u)-E_0(m)\right)\right)\;.
\label{p'phi}
\end{equation}
Note that, whereas in Einstein's dynamics ($p=Eu/c^2$, $p_0=0$, $\mu=0$) the right-hand side of Eq.~(\ref{p'phi}) turns out to be proportional to $E(m,u)$, this is not the case in general.  

Let us restrict ourselves, from now on, to the anisotropic dynamics of Sec.~\ref{subsec:kneqc}.  It is convenient to define the function
\begin{equation}
\Gamma(u):=\ee^{-h(u)/\kappa}\,\varphi(u)^{-1/2}\;.
\label{Gamma}
\end{equation}
Then, the momentum and energy in Eqs.~(\ref{pEinst}) and (\ref{EEinst}) can be rewritten as:
\begin{equation}
p(m,u)=\frac{m}{1/c^2-1/\kappa^2}\left(\mu+\frac{1-\varepsilon^2}{c^2}\,u\right)\Gamma(u)+p_0(m)-\frac{\mu\,m}{1/c^2-1/\kappa^2}\;;
\label{pEinstGamma}
\end{equation}
\begin{equation}
E(m,u)=\frac{m}{1/c^2-1/\kappa^2}\left(1+\left(\mu+2\varepsilon/c\right)u\right)\Gamma(u)+E_0(m)-\frac{m}{1/c^2-1/\kappa^2}\;.
\label{EEinstGamma}
\end{equation}
It is now easy, remembering Eq.~(\ref{muu}), to see that $p'(m,u)\varphi(u)=m\,\Gamma(u)$, so
\begin{equation}
M=m_1\,\Gamma(u_1)+m_2\,\Gamma(u_2)\;.
\label{M-new}
\end{equation}

Assume now that Eq.~(\ref{p0}) holds.  It is then easy to see that Eq.~(\ref{p+p}) amounts to 
\begin{equation}
m_1\,\Gamma(u_1)\,u_1+m_2\,\Gamma(u_2)\,u_2=0\;.
\label{mGammau}
\end{equation}
Using this result in computing the total energy of the system in the centre-of-momentum reference frame $\cal K$, and assuming also the validity of Eq.~(\ref{Emc2}), one finds 
\begin{equation}
E(m_1,u_1)+E(m_2,u_2)=\frac{M}{1/c^2-1/\kappa^2}\;.
\label{E+E}
\end{equation}
Hence, Eq.~(\ref{Emc2}) holds also for the system as a whole, not only for the individual particles.  Of course, this conclusion can be straightforwardly generalised to an arbitrary number of particles.

Note that this argument goes through only if $p_0$ and $E_0$ are of the form given by Eqs.~(\ref{p0}) and (\ref{Emc2}).  In fact, although the result concerns the notion of rest energy, its validity requires also that there be a rest momentum $p_0=\mu\,E_0$.  This is not an option: If one accepts the existence of a rest energy, such a rest momentum must also exist, for consistency, as can be seen by the following argument.  In the centre-of-mass reference frame $\cal K$, 
\begin{equation}
E(m_1,u_1)+E(m_2,u_2)=E_0(M)\;.
\label{EEM}
\end{equation}
In the arbitrary frame $\overline{\cal K}$,
\begin{equation}
E(m_1,\Phi(u_1,v))+E(m_2,\Phi(u_2,v))=E(M,v)\;.
\label{EEM-bar}
\end{equation}
Differentiating Eq.~(\ref{EEM-bar}) and setting $v=0$, we find 
\begin{equation}
E'(m_1,u_1)\varphi(u_1)+E'(m_2,u_2)\varphi(u_2)=E'(M,0)=0\;,
\label{E'E'}
\end{equation}
where the right-hand side can be evaluated, {\em e.g.\/}, using Eq.~(\ref{dT}) and the fact that $p'(M,0)=M$.  Using now Eq.~(\ref{momentum}) with $\lambda=1$, Eqs.~(\ref{nu}) and (\ref{beta'}), and finally Eqs.~(\ref{p+p}) and (\ref{EEM}), Eq.~(\ref{E'E'}) becomes
\begin{equation}
p_0(M)-\mu\,E_0(M)-p_0(m_1)-p_0(m_2)+\mu\,E_0(m_1)+\mu\,E_0(m_2)=0\;.
\label{pEpE}
\end{equation}
Remembering now that both $p_0$ and $E_0$ must be proportional to the mass (see end of Sec.~\ref{subsec:mass}), setting $p_0=\sigma\,E_0$, and using Eq.~(\ref{M-new}), we find easily that $\sigma=\mu$.

\subsection{Solving a partial differential equation}
\label{subsec:appendixC}
\renewcommand{\theequation}{\thesubsection.\arabic{equation}}
\setcounter{equation}{0}

Consider the first order linear partial differential equation
\begin{equation}
\left(AX+BY\right)\frac{\partial F}{\partial X}
+\left(CX+DY\right)\frac{\partial F}{\partial Y}=0\;,
\label{pde}
\end{equation}
where $A$, $B$, $C$, $D$ are constants.\footnote{Not to be confused with the four coefficients in Eq.~(\ref{trans}).}  Defining the new variable $Z:=Y/X$ and the function $G(X,Z):=F(X,XZ)$, Eq.~(\ref{pde}) can be rewritten as
\begin{equation}
\frac{1}{2}\,X\,\frac{\partial G}{\partial
X}+\frac{\Psi(Z)}{-\Psi'(Z)+\left(A+D\right)}\frac{\partial
G}{\partial Z}=0\;,
\label{pde1}
\end{equation}
where
\begin{equation}
\Psi(Z):=-BZ^2-\left(A-D\right)Z+C\;.
\label{psi}
\end{equation}
Defining
\begin{equation}
\Theta(Z):=\int_0^Z\frac{\d Z'}{\Psi(Z')}
\label{Theta}
\end{equation}
and a new variable $W$ through
\begin{equation}
\d W:=\frac{\d \Psi}{\Psi}-\left(A+D\right)\d\Theta\;,
\label{W}
\end{equation}
Eq.~(\ref{pde1}) can be immediately rewritten in the form
\begin{equation}
\frac{1}{2}\,X\,\frac{\partial G}{\partial X}-\frac{\partial
G}{\partial W}=0\;,
\label{pde2}
\end{equation}
where $G$ now denotes, with a little abuse of notation, the function $G$ when $Z$ is expressed in terms of $W$.  It is evident that $G$ can be any arbitrary function of $\ln X^2+W$, so $F$ is an arbitrary function of
\begin{equation}
X^2\Psi(Y/X)\ee^{-\left(A+D\right)\Theta(Y/X)}\;.
\label{solution}
\end{equation}
%

\end{document}